\documentclass[12pt,a4paper]{article}
\usepackage[usenames,dvipsnames]{pstricks} % latexdraw
\usepackage{afterpage}
\usepackage{epsfig}
\usepackage{pst-grad} % For gradients
\usepackage{pst-plot} % For axes
\usepackage[a4paper]{geometry}
\usepackage{float}
\usepackage[stable]{footmisc}
\usepackage[utf8x]{inputenc}
\usepackage{a4wide}
\usepackage{amsmath,amssymb,amstext,amsthm,amssymb}
\usepackage{amsfonts}
\usepackage{framed}
\usepackage{graphicx}
\usepackage{color}
\usepackage{bbold}
\usepackage{bbm}
\usepackage{rotating} %% rotating figures
\usepackage{slashed} %%feynman slash
\usepackage{cancel} %% diagonal line through word
\usepackage{braket} %% bra ket
\usepackage{amstext}
\usepackage{array}
\usepackage{setspace}
\usepackage{hyperref}
\usepackage{overpic}
\usepackage{bbm}
\usepackage{cite}
\usepackage{lipsum}
\newcommand{\del}[0]{\partial}
\let\baraccent=\=
\renewcommand{\=}[1]{\stackrel{#1}{=}}
\newcommand{\id}[0]{\mathbb{I}}
\newcommand{\Tr}[1]{\text{Tr}\left( #1 \right)}
\DeclareSymbolFontAlphabet{\mathbb}{AMSb}

\begin{document}

	\pagestyle{plain}

	%----------------------------------------------------------------------%
	%  numbering equations with section number
	%----------------------------------------------------------------------%
	\makeatletter
	\@addtoreset{equation}{section}
	\makeatother
	\renewcommand{\theequation}{\thesection.\arabic{equation}}
	%----------------------------------------------------------------------%
	%  title page
	%----------------------------------------------------------------------%
	\pagestyle{empty}
	
	\vspace{0.5cm}
	 
	\begin{center}
		{\Large \bf{Resolving spacetime singularities in flux compactifications \& KKLT}
			\\[15mm]}
		\normalsize{Federico Carta$^{1}$ and Jakob Moritz$^{2}$  \\[6mm]}
		\small{\slshape
			$^{1}$Department of Mathematical Sciences, Durham University, Durham DH1 3LE, UK\\
			$^{2}$Department of Physics, Cornell University, Ithaca, NY 14853, USA
			\\[8mm]  }
		\normalsize{\bf Abstract} \\[8mm]
	\end{center}
	\begin{center}
		\begin{minipage}[h]{15.0cm}
			
			In flux compactifications of type IIB string theory with D3 and seven-branes, the negative induced D3 charge localized on seven-branes leads to an apparently pathological profile of the metric sufficiently close to the source. With the volume modulus stabilized in a KKLT de Sitter vacuum this pathological region takes over a significant part of the entire compactification, threatening to spoil the KKLT effective field theory. In this paper we employ the Seiberg-Witten solution of pure $SU(N)$ super Yang-Mills theory to argue that wrapped seven-branes can be thought of as bound states of more microscopic exotic branes. We argue that the low-energy worldvolume dynamics of a stack of $n$ such exotic branes is given by the  $(A_1,A_{n-1})$ Argyres-Douglas theory. Moreover, the splitting of the perturbative (in $\alpha'$) seven-brane into its constituent branes at the non-perturbative level resolves the apparently pathological region close to the seven-brane and replaces it with a region of $\mathcal{O}(1)$ Einstein frame volume. While this region generically takes up an $\mathcal{O}(1)$ fraction of the compactification in a KKLT de Sitter vacuum we argue that a small flux superpotential \textit{dynamically} ensures that the 4d effective field theory of KKLT remains valid nevertheless.

		\end{minipage}
	\end{center}
	\newpage
	%----------------------------------------------------------------------%
	%  Resetting of counters
	%----------------------------------------------------------------------%
	\setcounter{page}{1}
	\pagestyle{plain}
	\renewcommand{\thefootnote}{\arabic{footnote}}
	\setcounter{footnote}{0}
	%----------------------------------------------------------------------%
	%  Paper begins
	%----------------------------------------------------------------------%

	\tableofcontents
	
	\newpage
	
	\section{Introduction}
	
	A promising and conceptually simple avenue towards understanding the nature of dark energy in quantum gravity is the pursuit of explicit de Sitter solutions in string theory \cite{Silverstein:2001xn,Maloney:2002rr,Kachru:2003aw,Kallosh:2004yh,Westphal:2006tn,Silverstein:2007ac,Dong:2010pm,Rummel:2011cd,Cicoli:2012vw,Louis:2012nb,Cicoli:2013cha,Marsh:2014nla,Braun:2015pza,Cicoli:2015ylx,Gallego:2017dvd}. It was argued long ago by Kachru, Kallosh, Linde and Trivedi (KKLT) that such solutions should exist in the landscape of type IIB Calabi-Yau (CY) orientifolds with threeform fluxes, D3 and seven-branes \cite{Kachru:2003aw}. 
	Specifically, KKLT showed that if the classical flux superpotential $W_0$ can be tuned to very small values, volume moduli are generically stabilized at large values, and the resulting vacuum is $AdS_4$. If moreover, the very same fluxes generate strongly warped and weakly curved Klebanov-Strassler throats \cite{Klebanov:2000hb,Giddings:2001yu}, an anti-D3 brane at the bottom of the throat can lift the vacuum energy to small positive values, provided that $W_0$ can be finely tuned against the throat hierarchy.

	Various aspects of the KKLT proposal have been scrutinized extensively in the past: some of the earliest work has focused on establishing that all K\"ahler moduli can in principle be stabilized non-perturbatively \cite{Giryavets:2003vd,Denef:2004dm,Denef:2005mm}. Only more recently, it was shown that flux vacua with exponentially small flux superpotential can be found systematically \cite{Demirtas:2019sip}, and furthermore that the required solutions hosting \textit{both} long warped throats \textit{and} small flux superpotentials simultaneously can likewise be found \cite{Demirtas:2020ffz,Blumenhagen:2020ire}. 
	Another part of the recent literature has focused on the consistency and meta-stability of the warped anti-D3-brane \cite{Bena:2009xk,Bena:2011hz,Bena:2011wh,Bena:2012bk,Bena:2012vz,Bena:2014jaa,Michel:2014lva,Cohen-Maldonado:2015ssa,Polchinski:2015bea,Cohen-Maldonado:2015lyb}, and its surrounding throat region \cite{Bena:2018fqc,Blumenhagen:2019qcg,Bena:2019sxm,Dudas:2019pls}. In particular, \cite{Cohen-Maldonado:2015ssa} explains nicely how the apparently pathological near-brane behavior of supergravity fields is resolved by brane polarization. In yet another line of work the 10d uplift of gaugino condensation effects in the 4d EFT was studied \cite{Baumann:2006th,Baumann:2007ah,Koerber:2007xk,Dymarsky:2010mf,Moritz:2017xto,Gautason:2018gln,Hamada:2018qef,Kallosh:2019oxv,Hamada:2019ack,Carta:2019rhx,Gautason:2019jwq,Kachru:2019dvo}. Specifically, after proper inclusion of quartic gaugino terms and contributions to the stress-tensor from the (non-local) volume dependence of the gaugino bi-linear, one arrives at a consistent picture \cite{Hamada:2018qef,Kallosh:2019oxv,Hamada:2019ack,Carta:2019rhx,Kachru:2019dvo} (see however \cite{Gautason:2019jwq}).
	
	In contrast, the aim of this paper is to resolve (literally) a parametric control problem as posed in \cite{Carta:2019rhx,Gao:2020xqh} (see also \cite{Freivogel:2008wm})  which we summarize in Section \ref{sec:fitting_throats}. Briefly stated, the problem is that an efficient competition between stabilizing potential and a meta-stable anti-D3 uplift forces the overall volume modulus to take values of the same order as the overall D3-charge hosted in the warped throat. In such a regime, the backreaction radii of both the sources of positive and negative D3 charges turn out to be of the same order as the typical length scale of the compact CY. 
	Since the square of the conformal factor of the compact threefold is driven to unbounded negative values near \textit{localized} sources of negative D3 charge with real co-dimension larger than one (see Section \ref{sec:classical-singularities}) this implies that at the KKLT minimum an entire $\mathcal{O}(1)$ fraction of the compact threefold can no longer be described by a semi-classical solution. This is to be contrasted with the large volume regime where only the regions very close to localized sources of negative D3 charge cannot be described by the semi-classical solution which is of course entirely expected, and irrelevant for the Physics at length-scales larger than the string scale.
	
	The aim of this paper is to understand in detail how string theory resolves this semi-classical singularity via effects that are non-perturbative in the $\alpha'$ expansion. In order to make progress, we will make the important simplifying assumption that all negative D3 charge is hosted on seven-branes wrapping K3 surfaces. While such orientifold vacua are not generic, many exist (see e.g. \cite{Carta:2020ohw}). Then, since the low-energy world-volume dynamics of the seven-branes accidentally respects $\mathcal{N}=2$ SUSY in four dimensions this will allow us to understand the relevant non-perturbative bulk effects in terms of instanton effects in the gauge theory. These in turn are known by virtue of the Seiberg-Witten solution of the low-energy limit of $\mathcal{N}=2$ pure Yang-Mills theory \cite{Seiberg:1994rs,Seiberg:1994aj,Klemm:1994qs,Argyres:1994xh,Klemm:1995wp}. Using this, we will argue for the following proposals:
	\begin{enumerate}
		\item \textit{The wrapped seven-branes hosting negative D3 charge must properly be thought of as bound states of more elementary co-dimension two exotic branes of $6d$ $\mathcal{N}=(2,0)$ SUGRA separated by a distance that is non-perturbatively small in the $\alpha'$-expansion.}
		\item \textit{As the overall volume modulus takes values of order of the D3 tadpole, the exotic branes forming the perturbative defects get split over a distance comparable to the size of the compact threefold. This splitting of branes stops the running of the conformal factor before it turns negative, thus resolving the apparent pathology.}
		\item \textit{The 'inside region' left behind by the spreading of exotic defects has $\mathcal{O}(1)$ Einstein frame curvature.}
	\end{enumerate}
	More precisely, we will resolve the singularity of the conformal factor much in the same way that F-theory \cite{Vafa:1996xn,Morrison:1996na,Morrison:1996pp} resolves the apparent singularity of the axio-dilaton in the vicinity of an O7 plane (where formally $1/g_s<0$) by splitting the perturbative O7 plane into a pair of mutually non-local defects \cite{Sen:1996vd,Banks:1996nj}.\footnote{Note that this is also morally similar to the resolution of the anti-D3 brane singularity at the bottom of a warped throat by polarization into a spherical NS5-brane as discussed in \cite{Cohen-Maldonado:2015ssa}.} This well-known effect can be seen either \textit{directly} using F-theory \cite{Sen:1996vd} or alternatively by probing the background with a D3 brane realizing the $SU(2)$ Yang-Mills theory as in \cite{Banks:1996nj}. Similarly, we can realize the $SU(N)$ Yang-Mills theory by wrapping $N$ D7 branes on K3, thus giving rise to $-N$ units of induced D3 charge \cite{Douglas:1996du}. We argue that the non-perturbative gauge theory effects of the Yang-Mills theory can directly be interpreted as a splitting of each perturbative D7 brane into a pair of mutually non-local exotic branes on which D5-strings can end.\footnote{As in \cite{Douglas:1996du} the stretched D5 string realizes a magnetic monopole.} By comparing with the known singularities in the moduli space of $SU(N)$ Yang-Mills theory \cite{Klemm:1994qs,Argyres:1994xh,Klemm:1995wp,Argyres:1995jj} we conclude that a stack of $n$ such exotic branes hosts the Argyres-Douglas (AD) theory of type $(A_1,A_{n-1})$. Thus, if we assume that a significant fraction of the D3 tadpole resides on a stack of seven-branes wrapping a K3 surface, we can establish our claim (a).
	
	Using (a) it follows immediately that the exotic branes are separated by a distance of order the strong coupling scale $\Lambda$ (in appropriate units). Thus, as the UV gauge coupling (the K\"ahler modulus) is driven towards stronger coupling, the exotic branes spread outwards in their transverse space. Since the UV coupling is naturally defined at distances of order the size of the compact threefold our claim (b) follows. Furthermore, in this regime the strong coupling scale $\Lambda$ is of the same order as R-symmetry breaking spurions $\Lambda_{\slashed{R}}$ induced by the compact bulk threefold. Therefore, the above regime is naturally associated with poor control over the low energy expansion in $\Lambda/\Lambda_{\slashed{R}}$.\footnote{Note that sub-leading corrections in $\Lambda/\Lambda_{\slashed{R}}$ are expected to respect only four supercharges.}
	
	Furthermore, we will use the Seiberg-Witten solution of $SU(N)$ Yang-Mills theory to argue that the 'inside region' is indeed strongly curved, i.e. (c). More precisely, we will show that the origin of the Coulomb branch of the gauge theory approaches a wall of marginal stability in the large $N$ limit. We will conclude from this that there is no (parametrically) preferred string with small tension that can be stretched in the interior region and thus Einstein frame volumes must be of $\mathcal{O}(1)$ in the large $N$ limit.
	
	As the resolution of semi-classical singularities can be understood entirely in terms of gauge theory effects known to be present in the 4d EFT we view this as strong evidence that the EFT employed by KKLT remains well-controlled even though an $\mathcal{O}(1)$ fraction of the bulk CY has $\mathcal{O}(1)$ Einstein frame curvature.
	More precisely, in Sections \ref{secsub:comments_throat_recombination} and \ref{sec:instanton_exp_KKLT} we will argue that
	\begin{enumerate}
		\setcounter{enumi}{3}
		\item \textit{A small classical flux superpotential ensures control over the euclidean D3 (ED3) instanton expansion even if an $\mathcal{O}(1)$ fraction of the ED3 worldvolume probes the strongly curved region.}
		\item \textit{A small classical flux superpotential also ensures that the non-perturbative spreading of defects that ensures the resolution of singularities does not extend far enough to significantly affect the local warped throat region.}
	\end{enumerate}
	We argue for (d) by noting that if the leading ED3 instanton wrapped on some divisor $D$ has disappeared entirely into the strongly curved inside region its euclidean action must be of $\mathcal{O}(1)$. Since a small $W_0$ ensures that its action is of order $\log(|W_0|^{-1})\gg 1$ we conclude that part of $D$ remains in the weakly curved outside region. Furthermore, we argue that if a significant fraction of positive D3 charge is hosted in a warped throat, the final stages of the small volume limit correspond to the exotic branes 'walking' down the warped throat leading to recombination of D3 charge (see Section \ref{secsub:comments_throat_recombination}). Thus, if $D$ does not reach all the way to the bottom of the throat, a large ED3 action dynamically ensures that the warped throat is not significantly affected by the strongly coupled effects in the bulk, i.e. (e).

	Since the KKLT proposal only requires 1) tunable control over the instanton expansion and 2) a tunable uplift we conclude that the presence of even a large region of $\mathcal{O}(1)$ curvature in the bulk should not jeopardize the EFT employed for moduli stablization and uplift. Nevertheless, our analysis suggests (as in \cite{Carta:2019rhx,Gao:2020xqh}) that a considerable fraction of the bulk CY has $\mathcal{O}(1)$ Einstein frame curvature in KKLT de Sitter vacua. This should be taken into account in phenomenological model building based on the KKLT proposal.
	
	This paper is organized as follows: We start by reviewing basics about the overall volume modulus and the local  behavior of the conformal factor near sources of D3 charge (Section \ref{sec:classical-singularities}). In Section \ref{sec:fitting_throats} we review the potential control problem of KKLT discussed in \cite{Carta:2019rhx,Gao:2020xqh}, related to fitting throats into compact bulk CYs and avoiding large singular regions in the bulk. In Section \ref{sec:resolution} we explain how non-perturbative effects cure the apparent singularities via the splitting of perturbative D7 branes into elementary monodromy defects of $6d$ $\mathcal{N}=(2,0)$ SUGRA bound together by a potential. In Section \ref{sec:comments} we comment on the nature of the small volume limit in a global model \ref{secsub:comments_global_model}, and relate its final stages with a process of recombination of D3 charges \ref{secsub:comments_global_model}, \ref{secsub:comments_throat_recombination}. Finally, in Section \ref{sec:instanton_exp_KKLT} we argue that KKLT de Sitter vacua are generically safe from significant alterations of the EFT due to the appearance of regions of strong bulk curvature. We conclude with Section \ref{sec:conclusions}.

\section{Spacetime singularities in flux compactifications}
\label{sec:classical-singularities}

As a starting point we consider the (classical) GKP solutions \cite{Giddings:2001yu}: tree level vacua of type IIB string theory on an O3/O7 orientifold $B$ of a Calabi-Yau (CY) threefold $X$ with D7 branes, D3 branes and threeform fluxes, or more generally F-theory on an elliptically fibered CY fourfold with base $B$ and four-form fluxes.\footnote{We work in 10d Einstein frame, and set the 10d reduced Planck mass $M_{P,10d}^8 =4\pi$. This amounts to setting $l_s^2\equiv (2\pi)^2\alpha'=1$. Moreover, we use our freedom to Weyl-rescale the 4d components of the 10d metric to also set the 4d reduced Planck mass to $M_{P,4d}^2=4\pi$.} Let $\{D_i\}$ be a basis of $H_4^+(X,\mathbb{Z})=H_{2,2}^+(X,\mathbb{Z})$, i.e. the orientifold-even four-cycles. Then, at the classical level, the $h^{1,1}_+(X)$ K\"ahler moduli $T_i$
remain exactly flat directions, while the $h^{2,1}_-$ complex structure moduli $z^a$ and the dilaton $\tau$ are frozen by the fluxes. Below the mass-scale of the K\"ahler moduli the classical (in the $\alpha'$ expansion) effective $\mathcal{N}=1$ superpotential is a constant $W_0$,
\begin{equation}
W_{cl}(T^i)=W_0:=\sqrt{\frac{2}{\pi}}\left\langle\int_{X}(F_3-\tau H_3)\wedge \Omega(z)\right\rangle+\mathcal{O}(e^{2\pi i\langle\tau\rangle})\, ,
\end{equation}
where the first term is the Gukov-Vafa-Witten (GVW) flux superpotential \cite{Gukov:1999ya}, and the $\mathcal{O}(e^{2\pi i \tau})$ $D(-1)$ instanton corrections can be computed in F-theory. 

As the discrete axionic shift symmetries $T^i\rightarrow T^i+ic^i$, $c^i\in \mathbb{Z}$, cannot be broken by perturbative effects, the superpotential receives only non-perturbative corrections in the K\"ahler moduli \cite{Witten:1996bn}
\begin{equation}\label{eq:np_superpotential}
W(T^i)=W_{cl}(T^i)+W_{np}(T^i)\equiv W_0+\sum_{\vec{n}\in \Gamma}A_{\vec{n}}\, e^{-2\pi\, \sum_i n_i T^i}\, ,
\end{equation}
where the $n_i$ are rational numbers, and the coefficients $A_{\vec{n}}$ are in general hard to compute. 

The 10d metric takes the form \cite{Giddings:2001yu}
\begin{equation}
ds^2=\frac{e^{2A(y)}}{t}dx²+e^{-2A(y)}g_{ij}^{\text{int}}dy^idy^j\, ,
\end{equation} 
with \textit{warp factor} $e^{2A(y)}$ varying over the 6d internal Calabi-Yau orientifold $B$ (or more generally an F-theory base) parameterized by local coordinates $y^i$. The metric $g_{ij}^{\text{int}}$ is a Ricci flat metric (or more generally an F-theory solution) normalized to unit volume, 
\begin{equation}
\int_{B}d^6y\sqrt{g^{\text{int}}}=1\, ,
\end{equation}
and $dx^2$ denotes the 4d Minkowski line element. The \textit{overall volume modulus} $t$ is defined as \cite{Giddings:2005ff}
\begin{equation}\label{eq:overall_vol_modulus}
t:=\int_{B}d^6y\sqrt{g^{\text{int}}}e^{-4A(y)}\, .
\end{equation}
The \textit{physical} metric of $B$ differs from $g_{ij}^{\text{int}}$ by the crucial factor $e^{-2A}$ which by virtue of the equations of motion is the inverse of the warp factor. We will refer to its square $e^{-4A}$ as the \textit{conformal factor}. 

In terms of the 10d metric and four form potential $C_4$ the K\"ahler moduli are defined as
\begin{equation}\label{eq:def_Kahler_moduli}
T^i:=\int_{D_i} e^{-4A}\frac{1}{2}J\wedge J-iC_4\, ,
\end{equation}
where $J$ is the K\"ahler form of the unit-volume orientifold or F-theory base $B$.

A GKP solution \cite{Giddings:2001yu} is given by a choice of flux quanta $[F_3],[H_3]\in H^3(X,\mathbb{Z})$ stabilizing the complex structure moduli of $B$ and the axio-dilaton $\tau:=C_0+ie^{-\phi}$ at values such that
\begin{equation}
*_{\text{int}}G_3=i G_3\, ,\quad G_3:=F_3-\tau H_3\, .
\end{equation}
The self-dual five form field strength $F_5$ becomes
\begin{equation}
F_5=(1+*)d e^{4A}\wedge d^4x\, ,
\end{equation}
and the conformal factor is a solution to the electro-static problem
\begin{equation}\label{eq:electro-static-eq}
-\nabla^2_{\text{int}}e^{-4A}=\rho_{D3}\, ,
\end{equation}
with D3 charge density $\rho_{D3}$ with overall net zero charge, and $\nabla^2_{\text{int}}$ is the Laplacian associated with the metric $g_{ij}^{\text{int}}$. The fluxes carry positive D3 charge density $F_3\wedge H_3$, mobile D3 branes and O3 planes have localized charge $+1$ respectively $-\frac{1}{4}$, while seven-branes wrapped on a surface $S$ carry negative D3-charge smeared across their world-volume proportional to the Euler characteristic $\chi(S)$.

Eq. \eqref{eq:electro-static-eq} clearly has a modulus $e^{-4A(y)}\rightarrow e^{-4A(y)}+\delta t$ corresponding to shifts of the overall volume modulus $t\rightarrow t+\delta t$ in eq. \eqref{eq:overall_vol_modulus}. The reason for this terminology is the fact that in the large $t$ limit we get $e^{-4A(y)}\rightarrow t=const$ (except near singularities) so the 10d metric approaches the form
\begin{equation}
ds^2_{t\rightarrow \infty}=t^{-\frac{3}{2}}dx^2+t^{\frac{1}{2}}g_{ij}^{\text{int}}dy^i dy^j\, .
\end{equation}
Therefore, in this limit the physical volume of $B$ becomes $\text{Vol}(B)\rightarrow t^{\frac{3}{2}}\gg 1$. In the vicinity of singular sources the conformal factor behaves as it would in flat space: near a stack of $N$ D3 branes we get
\begin{equation}
e^{-4A(r)}=\frac{N}{4\pi^3r^4}+c\, ,
\end{equation}
where $r$ is a radial distance measuring radial distance from the source in the unit volume metric $g_{ij}^{\text{int}}$, and $c$ is an integration constant. Shifts $c\rightarrow c+\delta c$ correspond to shifts $t\rightarrow t+\delta c$, and at distances much smaller than the compactification scale the integration constant $c$ is meaningless because it can be absorbed by a redefinition of the radial coordinate:
\begin{equation}
\sqrt{\frac{N}{4\pi^3r^4}+c}(dr^2+r^2 d\Omega_{S^5})\equiv \sqrt{\frac{N}{4\pi^3\tilde{r}^4}+1}(d\tilde{r}^2+\tilde{r}^2 d\Omega_{S^5})\, ,
\end{equation}
with $\tilde{r}^4:=c \, r^4$.

Near a (real) co-dimension two source wrapped on a surface $S$ the conformal factor scales as
\begin{equation}
e^{-4A(r)}=-\frac{\rho_{D3,\perp}}{2\pi}\log(r/r^*)\, ,
\end{equation}
where $\rho_{D3,\perp}$ is the (assumed constant) D3-charge density along the surface $S$, and
with \textit{dynamically generated} radial scale $r^*$. Rescalings $r^*\rightarrow r^*e^{\frac{2\pi}{\rho_{D3,\perp}} \delta t }$ correspond to shifts in the volume modulus, $t\rightarrow t+\delta t$. Thus, for \textit{negative} D3 charge density, and for large volume $t$, one formally gets a \textit{negative} conformal factor below an exponentially small radial distance 
\begin{equation}
r\leq r^*\sim e^{-\frac{2\pi t}{|\rho_{D3,\perp}|}}=e^{-\frac{2\pi}{|Q|}\text{Vol}(S)}\, ,
\end{equation}
where $Q$ is the integrated D3 charge on the surface $S$. This formula of course already suggests that the negative conformal factor regime is cured by \textit{non-perturbative} effects in the $\alpha'$-expansion, which will indeed turn out to be true.

Now, as in (the appendix of) \cite{Carta:2019rhx}, let us consider the case where all positive D3 charge is localized at real co-dimension six (a stack of $N$ D3 branes), and all negative D3 charge is localized on a complex surface $S$ at real co-dimension two (a seven-brane stack with induced D3 charge $-N$). As usual, by inspecting the D3 stack one sees that its backreaction turns it into an $AdS_5\times S^5$ throat of radius $N^{1/4}$. This throat can be glued into a weakly curved bulk only when the physical bulk-volume  is large enough,
\begin{equation}\label{eq:fitting_charge}
\text{Vol}(B)^{\frac{2}{3}}\gtrsim N\, .
\end{equation}
This geometrical argument can be verified at the technical level via inspection of the solution near the seven-branes: in the regime $\text{Vol}(S)\gg |Q|=N$, where the throat is parametrically smaller than the bulk, the conformal factor is negative only exponentially close to the brane stack. In contrast, in the critical regime $\text{Vol}(S)\sim N$, where the throat fits into the bulk only marginally, the exponential suppression gets lost and the vanishing locus of the conformal factor has moved into generic position \cite{Carta:2019rhx,Gao:2020xqh}. This latter point has been emphasized and discussed in detail particularly in \cite{Gao:2020xqh}.\footnote{It is not essential to the argument what the general charge configuration is as long as we keep negative and positive D3 charge seperated from each other, to avoid recombination or significant screening of brane charges. A detailed discussion of this can be found in \cite{Gao:2020xqh}.}

To our knowledge the regime where the conformal factor is formally negative in an $\mathcal{O}(1)$ fraction of the CY has remained largely unstudied so it has been standard practice to impose eq. \eqref{eq:fitting_charge} with some reasonable control factor \cite{McAllister:2008hb,Freivogel:2008wm,Carta:2019rhx}. Part of the aim of this paper is to relax this requirement and see where it takes us.

\section{Fitting throats \& singularities in KKLT}
\label{sec:fitting_throats}

In the KKLT scheme of moduli stabilization one stabilizes the K\"ahler modulus $T$ (we assume $h^{1,1}_+(X)=1$) using a small classical flux superpotential $W_0\ll 1$ and effects non-perturbative in the $\alpha'$ expansion $\sim e^{-\frac{2\pi}{\texttt{c}} \text{Re}(T) }$, where $\texttt{c}\in \mathbb{N}$ is the dual-Coxeter number of a confining seven-brane gauge theory or $\texttt{c}=1$ for a euclidean D3 brane (ED3) instanton wrapping the generator of $H_4(X,\mathbb{Z})$ \cite{Kachru:2003aw}. This leads to moduli stabilization at large volume
\begin{equation}
\text{Re}(T)\approx \frac{\texttt{c}}{2\pi}\log(|W_0|^{-1})\, ,
\end{equation}  
if the classical flux superpotential is small, $|W_0|\ll 1$, and the scalar potential is of order
\begin{equation}
V_{\text{bulk}}\sim -e^{-4\pi \text{Re}(T)/c}\sim -|W_0|^2<0\, .
\end{equation}
For simplicity, let us set $\texttt{c}=1$, so we consider moduli stabilization from euclidean D3 brane instantons. The negative scalar potential from the bulk moduli stabilization must be compensated for by a small `uplifting' potential energy from an anti-D3 brane at the bottom of a warped throat. Its contribution to the scalar potential is gravitationally redshifted by a factor \cite{Klebanov:2000hb,Giddings:2001yu,Kachru:2003aw}
\begin{equation}\label{eq:IR_warp_factor}
a_0^4:=\frac{e^{4A}|_{IR}}{e^{4A}|_{UV}}\sim e^{-\frac{8\pi}{3}\frac{K}{g_sM}}=e^{-\frac{8\pi}{3}\frac{Q_{\text{throat}}}{g_sM^2}}\ll 1\, ,
\end{equation}
where $K$ and $M$ are flux quanta, $g_s$ is the string coupling, and $Q_{\text{throat}}:=KM$ is the D3 charge of the fluxes generating the throat. By appropriate fine-tuning of the IR warp factor $a_0^4\approx |W_0|^2\sim e^{-4\pi \text{Re}(T)}$, i.e.
\begin{equation}\label{eq:dS_tuning}
g_sM^2\overset{!}{\sim}\frac{Q_{\text{throat}}}{\text{Re}(T)}\, ,
\end{equation}
one can find vacua with small positive cosmological constant \cite{Kachru:2003aw}. Furthermore, for the throat to fit into the bulk CY the r.h. side of \eqref{eq:dS_tuning} should be smaller than unity, so we obtain a constraint
\begin{equation}\label{eq:throatfitting}
\text{Re}(T)\gtrsim Q_{\text{throat}}\quad \rightarrow \quad g_sM^2\lesssim 1\, .
\end{equation}
However, for a confident prediction of the meta-stability of the anti-D3 brane \cite{Kachru:2002gs} (see also \cite{Bena:2019sxm,Blumenhagen:2019qcg}) one would like to stay at weak string coupling $g_s\ll 1$ and in the regime of validity of the 10d SUGRA approximation at the tip of the throat: $R^2_{IR}\sim g_sM\gg 1$, so
\begin{equation}\label{eq:SUGRAcontrol}
g_sM^2=\frac{(g_sM)^2}{g_s}\gg 1\, .
\end{equation}
The (parametric) tension between eq. \eqref{eq:SUGRAcontrol} and eq. \eqref{eq:throatfitting} suggests a difficulty to 'fit' an appropriately red-shifted and weakly curved throat into a likewise weakly curved bulk geometry with moduli stabilization in place, as was noted in \cite{Carta:2019rhx}.\footnote{Note that the problem becomes more severe when the uplift scale is decreased, i.e. for AdS vacua with small SUSY breaking, the prospects of which have recently been studied in more generality in \cite{Hebecker:2020ejb}. We note that the results of this paper may also be used to argue for the possibility of small SUSY breaking in AdS using KKLT.} Upon relaxing the condition \eqref{eq:throatfitting} one has to deal with the fact that the classical vanishing locus of the conformal factor $e^{-4A}$ is no longer exponentially close to the position of the seven-branes. In fact, the region where formally $e^{-4A}<0$ will generically take up an $\mathcal{O}(1)$ fraction of the CY \cite{Carta:2019rhx,Gao:2020xqh}.

\section{Resolution of singularities}
\label{sec:resolution}

In this section we would like to explain our proposal that seven-branes wrapped on K3 split into elementary monodromy defects of $6d$ $\mathcal{N}=(2,0)$ supergravity, argue that stacks of these host Argyres-Douglas SCFTs of type $(A_1,A_n)$, and explain how this resolves the apparent singularities of the conformal factor (see Section \ref{secsub:resolution_qualitative}). Furthermore, we will argue that after this is taken into account, the bulk region where semi-classically $e^{-4A}<0$ gets replaced by a region of $\mathcal{O}(1)$ Einstein frame curvature (in Section \ref{secsub:resolution_marginal_stability}).

But first, as a proof of principle, let us exhibit an $\mathcal{N}=1$ O7 orientifold of a CY threefold such that the entire D3 tadpole is generated by seven-branes wrapping K3 surfaces. Many such examples can be found in the list of \cite{Carta:2020ohw}, and here is a particularly simple example: We consider the anti-canonical hypersurface in $\mathbb{P}^1\times \mathbb{P}^3$ which has two K\"ahler moduli associated with the hyperplane classes of $\mathbb{P}^1$ and $\mathbb{P}^3$ (and it is a K3 fibration over $\mathbb{P}^1$). The orientifold involution inherited from the ambient space involution
\begin{equation}
\mathcal{I}:\, \mathbb{P}^1\times \mathbb{P}^3\rightarrow \mathbb{P}^1\times \mathbb{P}^3\, ,\quad ([x_0:x_1],[y_0:...:y_4])\mapsto ([-x_0:x_1],[y_0:...:y_4])
\end{equation}
gives rise to two non-intersecting O7 planes at the intersection of $\{x_0=0\}$ respectively $\{x_1=0\}$ with the hypersurface. Both arise as an anti-canonical hypersurface in $\mathbb{P}^3$ which is K3. Placing four D7 branes on each O7 plane gives rise to gauge algebra $so(8)^2$ (which can be enhanced to $so(16)$ by moving all eight seven branes onto the same O7 plane) and induced D3 charge $-12$ that can be canceled by fluxes and/or mobile D3 branes. In the following we will discuss the resolution of singularities of the conformal factor in terms of the Seiberg-Witten solution of $SU(N)$ Yang-Mills theory, but the discussion for gauge group $SO(2N)$ is analogous. 

\subsection{A qualitative description}
\label{secsub:resolution_qualitative}

We will start by making an analogy to F-theory.
The technical point that has to be addressed is the apparent singularity where the conformal factor turns negative in the proximity of seven-brane stacks with negative D3 charge,
\begin{equation}\label{eq:warp_factor_one_loop}
e^{-4A(r)}=\frac{|\rho_{D3}|}{2\pi}\log(r/r^*)<0\, ,\quad \text{for }r<r^*\, .
\end{equation}
It turns out that such apparent pathologies are completely analogous to the behavior of the dilaton near a \textit{perturbative} (in $g_s$) O7 plane:
\begin{equation}\label{eq:dilaton_one_loop}
e^{-\phi}(r)=\frac{4}{2\pi}\log(r/r^*)\, .
\end{equation}
Here, $e^{-\phi}$ is the radially dependent inverse string coupling, and the coefficient in front of the logarithm arises because an O7 plane has $-4$ units of D7 brane charge. At weak \textit{bulk} string coupling\footnote{To define this one uses the fact that the negative D7 charge of the O7 is canceled by other seven-branes screening the charge at large radii. At yet larger radii the dilaton profile is constant, and this constant is a modulus that we call $1/g_s$.} the singularity is exponentially close to the O-plane, i.e. at distances $\mathcal{O}(e^{-\frac{2\pi}{g_s}})$, but at $g_s\sim 1$ the negative region appears on a generic slice through the bulk. This singularity is of course resolved by D(-1) instanton effects, naturally described by F-theory: The O7 plane is a bound state of two $(p,q)$ seven-branes with monodromy charges \cite{Sen:1996vd}
\begin{equation}\label{eq:USp2_pq}
(2,-1) \quad \text{and} \quad  (0,1)\, ,
\end{equation}
separated from each other at non-perturbative distance $\sim e^{-\frac{2\pi}{g_s}}$.

Another way to describe the same phenomenom is to consider the setup from the point of view of a probe D3 brane. From this perspective, instanton effects in the gauge theory living on the D3 worldvolume will resolve the dilaton singularity, as first noted in \cite{Banks:1996nj}. Concretely, a probe D3 brane has gauge coupling $\frac{4\pi}{g^2}=\frac{1}{g_s}$ and in the vicinity of an O7 plane the $U(1)$ gauge group on its worldvolume enhances to $USp(2)\simeq SU(2)$. The position modulus of the D3 brane in the transverse space is identified with the vacuum expectation value (vev) of the  Coulomb branch (CB) operator of the worldvolume theory, which is $4$d $\mathcal{N}=2$ pure $SU(2)$. The low energy dynamics of this theory is determined by the Seiberg-Witten solution \cite{Seiberg:1994rs,Seiberg:1994aj}. In particular, in the interior of the CB there are \textit{two} monodromy defects  where dyons carrying electric-magnetic charges $(2,-1)$ respectively $(0,1)$ become massless. By identifying the W-boson as the stretched F-string, and the magnetic monopole as the D-string, one sees that the defects indeed correspond to the $(p,q)$ seven-branes listed in eq. \eqref{eq:USp2_pq}. Note that the D3 brane is a particularly \textit{clean} probe of the axio-dilaton profile: the gauge coupling corresponds to the axio-dilaton which is sourced \textit{only} by the background O7 plane and not by the probe itself.

Given the qualitative and even technical similarity of the apparent singularity of the conformal factor near loci of negative D3 charge \eqref{eq:warp_factor_one_loop} and the dilaton singularity \eqref{eq:dilaton_one_loop}, and given the well known resolution of the latter problem in terms of a probe D3 brane dynamics \cite{Banks:1996nj}, we find it natural and promising to consider very analogous reasoning in order to discuss the resolution of the former.

In order for this to work, one would like to relate the conformal factor $e^{-4A}$ to the Yang-Mills coupling of wrapped branes. Thus, we should consider a stack of $N$ wrapped D7 branes on a surface $S$. Upon moving away a single D7 brane in the transverse direction, its effective $U(1)$ coupling probes the backreaction of the remaining seven-branes. This deformation is a flat direction only if the normal bundle of the surface $S$ is trivial: classically, the Coulomb branch is explored by simply moving D7 branes away from the stack in the transverse directions. If the normal bundle $\mathcal{N}$ were non-trivial, there would \textit{either} not exist a Coulomb branch at all (the divisor is rigid) \textit{or} a D7 brane deformed away from the main stack still intersects it along a curve leading to massless bifundamental matter. Since $c_1(\mathcal{N})=-c_1(S)$ via the adjunction formula, we avoid this if we consider $N$ D7 branes wrapping a K3 surface in a CY(-orientifold) $X$. The holographic correspondence between the one-loop running of the gauge coupling and the solution for the conformal factor has been observed in \cite{Douglas:1996du} which also identifies the W-bosons as stretched F-strings and the magnetic monopoles as stretched effective strings from D5 branes wrapped on K3. Indeed, by evaluating the probe brane actions on the supergravity background sourced by the seven-branes, one recovers the field theory central charges of W-bosons and magnetic monopoles (see e.g. our Appendix \ref{app:bulk_eom}).

The precise holographic dictionary between the 7-brane gauge coupling and the conformal factor is\footnote{Strictly speaking, in eq. \eqref{eq:7brane_coupling} we should replace $\tau$ by the axio-dilaton averaged over the K3 surface.}
\begin{equation}\label{eq:7brane_coupling}
\tau_{D7}:= \hat{\tau}-\tau\, ,\quad \hat{\tau}:=\int_{K3}C_4+i \int_{K3}d^4y \sqrt{g^{\text{int}}_{K3}}e^{-4A}\, ,
\end{equation}
where $g^{\text{int}}_{K3}$ is the induced metric of the K3 surface obtained from the unit volume bulk metric $g^{\text{int}}$. The negative correction by the axio-dilaton $\tau$ is due to the $\mathcal{O}(\alpha'^2)$ curvature correction to the gauge-kinetic term living on the brane stack.\footnote{The correction by $\text{Re}(\tau)=C_0$ follows immediately from the well-known $\alpha'$-corrected D-brane CS action \cite{Green:1996dd,Cheung:1997az}. The correction by $\text{Im}(\tau)$ can most easily be seen from the fact that a gauge-instanton with action $2\pi\text{Im}(\tau_{D7})$ can be thought of as a wrapped euclidean D3 brane: the DBI action of a D3 brane receives a correction $\delta S=-\frac{1}{192\pi} \int_{K3} \text{Im}(\tau)\Tr{\mathcal{R}\wedge *\mathcal{R}}=-2\pi \text{Im}(\tau)$ \cite{Dasgupta:1997cd,Bachas:1999um,Fotopoulos:2001pt}.} Indeed, the log-coefficient of $\hat{\tau}-\tau$ is given by the difference of D7 and D3 brane charges $(Q_{D7},Q_{D3})=(N,-N)$, i.e.
\begin{equation}
\tau_{D7}(z)=-\frac{2N}{2\pi i}\log(z/z^*)+...\, ,
\end{equation}
where $z$ is the complex transverse coordinate. This matches with the beta-function coefficient $2N$ of $SU(N)$ Yang-Mills theory. Naturally, the distance scale $z^*$ corresponds to the strong coupling scale $\Lambda$ of the gauge theory. Thus, in order to describe what happens at $|z|<|z^*|$ one can again invoke the Seiberg Witten solution of pure $SU(N)$ $\mathcal{N}=2$ gauge theory. 

Ideally, one would like to determine the supergravity solution by probing $N$ background D7 branes with a further D7 brane that itself does not source the gauge coupling $\tau_{D7}$. However, unlike the D3 brane as a probe of a single O7-plane solution a probe D7 brane itself also sources $\tau_{D7}$. Therefore, one has to be a bit careful in identifying the Coulomb branch of the gauge theory with the physical transverse space of the brane stack. Happily, we will see that from the behavior of $SU(2)$ Yang-Mills theory one can quite readily understand how a single D7-brane splits into more elementary monodromy defects. The generalization to $SU(N)$ turns out to be straightforward.

Just as F-theory describes 'elementary' mutually non-local $SL(2,\mathbb{Z})$ monodromy defects in 10d, i.e. $(p,q)$-seven-branes, one should be able to describe the corresponding defects in our situation as monodromy defects\footnote{For an exposition of exotic defects in string theory, see e.g. \cite{deBoer:2012ma}} of 6d $\mathcal{N}=(2,0)$ SUGRA: type IIB string theory on K3 as in \cite{Martucci:2012jk,Braun:2013yla,Candelas:2014jma,Candelas:2014kma}. The U-duality group and spectrum of strings is much larger in 6d than it is in 10d \cite{Aspinwall:1996mn},
\begin{equation}
U=O(5,21;\mathbb{Z})\, ,
\end{equation}
and we get $\vec{p}$-strings with $\vec{p}\in \mathbb{Z}^{26}$ from 10d $(p,q)$ strings, $(p,q)$ 5-branes wrapped on K3, and D3 branes wrapped on the 22 two-cycles of K3, coupling to the (anti-)self-dual 2-forms in 6d transforming in the \textbf{26} of the monodromy group $U$. As in ten dimensions, there are cosmic defects on which (some) strings can end.\footnote{A cosmic defect is defined by an element of the U-duality group $M\in U$. A necessary condition for a $\vec{p}$-string to be able to end on it is that its tension is invariant under $M$.} We consider configurations of monodromy defects such that the monodromy transformations on overlapping patches are contained in a 
$U':=O(2,2;\mathbb{Z})\subset U$ subgroup acting only on the $(p,q)$ strings and wrapped $(p,q)$ five-branes transforming in the \textbf{4}, as in \cite{Braun:2013yla}. The tension of a $\vec{p}$-string, with $\vec{p}\in \mathbb{Z}^4$, in $6d$ Planck units is given by \cite{Braun:2018fdp}
\begin{equation}
\mathcal{T}_{\vec{p}}=\sqrt{\pi}\,\text{Im}(\tau)^{-\frac{1}{2}}\text{Im}(\hat{\tau})^{-\frac{1}{2}}|p_1\tau +p_2 \hat{\tau}+p_3-p_4\tau\hat{\tau}|\, .
\end{equation}
$U'$ is generated by two commuting $SL(2,\mathbb{Z})$ sub-groups acting on the modular parameters $(\tau,\hat{\tau})$, as well as a $\mathbb{Z}_2$ subgroup interchanging $\hat{\tau}\leftrightarrow \tau$ \cite{Braun:2013yla}. The latter can morally be thought of as 'four T-dualities' because the string-frame K3 volume gets inverted. Indeed, $\text{Im}(\tau)=C_0$ and $\text{Im}(\hat{\tau})=\int_{K3}C_4$ so the RR potentials transform according to the standard rules of T-dualities.

Configurations of cosmic defects whose monodromies are contained in $U'$ include in particular configurations of $(p,q)$ seven-branes wrapped on K3 and D3 branes on points in K3.
Since $O(2,2;\mathbb{Z})\subset O(2,18;\mathbb{Z})$ which is the monodromy group of the complex structure moduli space of a K3 surface, one may encode the axio-dilaton, conformal factor and $C_4$ profiles in a K3 surface fibered over the base $\mathbb{P}^1$ \cite{Martucci:2012jk,Braun:2013yla}. This approach has been termed \textit{G-theory} by analogy to F-theory \cite{Braun:2013yla}. As a consequence, each such brane configuration is dual to type IIB on a certain K3-fibered CY threefold.

We will now invoke the Seiberg-Witten solution of $SU(N)$ gauge theory living on a stack of $N$ D7 branes wrapped on K3, in order to understand the resolution of the singularity of the conformal factor, in the spirit of \cite{Banks:1996nj}. First, let us recall some well-known facts about the Seiberg-Witten solution. Pure $\mathcal{N}=2$ $SU(N)$ Yang-Mills theory has an $N-1$ dimensional Coulomb branch. At generic vev's of the adjoint scalar $\Phi$ the effective theory has gauge group $U(1)^{N-1}$ and an over-complete set of coordinates on moduli space is given by the \textit{periods} of an auxiliary Riemann surface $\Sigma_{N-1}$ of genus $N-1$. The latter can be taken to be the hypersurface in $\mathbb{C}^2\ni (x,y)$ specified by \cite{Seiberg:1994rs,Seiberg:1994aj,Klemm:1994qs,Klemm:1995wp,Argyres:1994xh} (we follow the conventions of \cite{Klemm:1995wp})
\begin{equation}\label{eq:SWcurve}
y^2=\mathcal{W}(x;\{u_k\})^2-\Lambda^{2N}\, ,
\end{equation}  
where $\Lambda$ is the strong coupling scale of the UV theory and $\mathcal{W}(x;\{u_k\})^2$ is the characteristic polynomial
\begin{equation}\label{eq:char_pol}
\mathcal{W}(x):=\det\left(x\id-\Phi\right)\equiv x^N-\sum_{k=2}^{N}u_k x^{N-k}\, ,
\end{equation}
and the \textit{Casimirs} $u_k$ are the invariant polynomials $u_k=\frac{1}{k}\Tr{\Phi^k}+\mathcal{O}(u_{k-1},...,u_2)$. One can think of the curve $\Sigma_{N-1}$ as the double cover of the complex $x$-plane with branch cuts running between the $N$ roots $e^+_i$ of $\mathcal{W}_+:=\mathcal{W}-\Lambda^N$ and the roots $e^-_i$ of $\mathcal{W}_-:=\mathcal{W}+\Lambda^N$.

One defines a holomorphic one-form (the \textit{Seiberg-Witten form} or \textit{differential})
\begin{equation}\label{eq:SWform}
\lambda(u):=\frac{1}{2\pi i}\frac{x\del_x \mathcal{W}}{y}dx\, ,
\end{equation}
and the period vector is defined as the integral of $\lambda$ over a basis of one-cycles. In the singular classical limit $\Lambda/u_k^{\frac{1}{k}}\rightarrow 0$ we get coinciding roots
\begin{equation}
e^{\pm}_i\rightarrow e_i\, ,
\end{equation}
where the $e^i$ are the roots of $\mathcal{W}(x)$. It follows that at weak coupling there is a preferred set of $N$ shrinking cycles $\alpha_i$ that encircle each coinciding pair of roots in the $x$-plane. In this limit the residue theorem implies that the associated period components are equal to the roots,
\begin{equation}
a_i:=\int_{\alpha_i}\lambda\longrightarrow e_i\, ,\quad i=1,...,N\, .
\end{equation}
We have $\sum_i a_i=0$ because $\sum_i [\alpha_i]=0\in H_1(\Sigma_{N-1},\mathbb{Z})$, so we can use $N-1$ of the $a_i$ as coordinates on moduli space in the weak coupling patch. In our context of realizing the gauge theory with $N$ D7 branes wrapped on K3 we can think of the roots $e_i$ as the positions of the $N$ \textit{perturbative} D7 branes in the transverse plane parameterized by $x\in \mathbb{C}$. Indeed, the charged W-bosons stretching between the $i$-th and $j$-th D7 brane have central charge
\begin{equation}
\mathcal{Z}_{ij}=a_i-a_j\, ,
\end{equation}
so the mass of the corresponding BPS particle is proportional to the distance between the branes in flat space parameterized by a flat coordinate $a$, which matches the mass of a stretched F-string (see Appendix \ref{app:bulk_eom}). Away from the classical limit each root $e_i$ splits in two,
\begin{equation}
e_i\rightarrow (e_+^i,e_-^i)\, ,
\end{equation}
separated from each other at a non-perturbative distance scale. Thus, even away from the classical limit we can think of the $a_i$'s as the center of mass positions of the $N$ D7 branes, but each D7 brane should be interpreted as a \textit{bound state} of a \textit{pair} of elementary monodromy defects at positions $(e_+^i,e_-^i)$. The two constituents are interpreted as elementary monodromy defects that are separated from each other by a non-perturbative distance scale in the transverse plane.

\begin{figure}
	\centering
	\includegraphics[keepaspectratio,width=12cm]{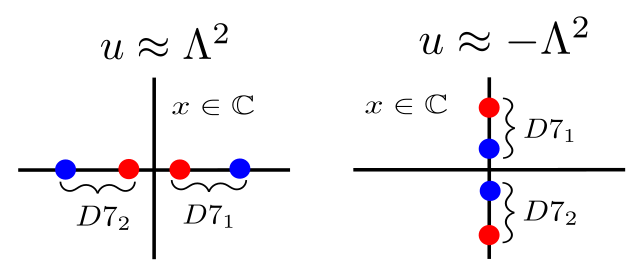}
	\caption{The roots of $\mathcal{W}_+(x)$ (blue) and $\mathcal{W}_-(x)$ in the complex $x$-plane. A pair of roots of $\mathcal{W}_+\mathcal{W}_-$ that coincides as $\Lambda\rightarrow 0$ can be thought of as a perturbative D7-brane. Left: roots near the first singularity $u\rightarrow +\Lambda^2$. Right: roots near the other singularity $u\rightarrow -\Lambda^2$.}
	\label{fig:D7_N=2}
\end{figure}

For simplicity, let us first consider the case $N=2$: for two D7 branes we have $\mathcal{W}(x)=x^2-u$ with $u:=\frac{1}{2}\Tr{\Phi^2}$ a gauge invariant coordinate on the Coulomb branch. As famously shown in \cite{Seiberg:1994rs} the classical singularity at $u=0$ is resolved into a pair of singularities at $u=\pm \Lambda$ where magnetic monopoles become massless. At each of these two points one of the two defects forming the first perturbative D7 brane coincides with one of the two defects forming the other one, see Figure \ref{fig:D7_N=2}. As shown in \cite{Douglas:1996du}, the magnetic monopole is a stretched string obtained from wrapping a D5 brane on K3. Since this is the dyon that becomes massless at one of the two singularities in moduli space 
we learn that a wrapped D5 string can end on one of the two elementary monodromy defects (or exotic branes) of 6d $\mathcal{N}=(2,0)$ supergravity that constitute a perturbative D7 brane. 

The monodromies of $(\tau,\hat{\tau})$ can be determined using G-theory. In fact, the appropriate K3 fibrations over $\mathbb{P}^1$ with monodromies only in $U'=\mathcal{O}(2,2;\mathbb{Z})$ have been described in \cite{Braun:2013yla}, and the relevant monodromy matrices have been worked out in \cite{Braun:2018fdp}, albeit in a very different context.\footnote{We are indebted to Andreas Braun for pointing out and explaining ref. \cite{Braun:2018fdp} to us. In an earlier version of this paper we had given the monodromies under the incorrect assumption that they act like a standard $SL(2,\mathbb{Z})$ transformation on $\tau_{D7}$.} Here, we present a simplified argument to get to the right result: first, one notices that the perturbative monodromy around a D7 brane acts as $\tau\rightarrow \tau+1$ and $\hat{\tau}\rightarrow \hat{\tau}-1$, in particular leaving the combination $\hat{\tau}+\tau$ invariant. Let us make the natural assumption that the microscopic monodromies also leave $\hat{\tau}+\tau$ invariant. Then, the only allowed monodromies are generated by $\{T^2,(-1)\}$ with $T:\,\tau_{D7}\mapsto \tau_{D7}+1$, and $(-1):\,\tau_{D7}\mapsto -\tau_{D7}$. Using that the D5 string (with tension $\propto |\tau_{D7}|$) can end on the first defect, as in \cite{Douglas:1996du}, it follows that
\begin{equation}\label{eq:monodromy1}
M_1=(-1)\, : \, \tau_{D7}\mapsto -\tau_{D7}\, ,\quad M_2=(-1)\cdot T^{-2}: \, \tau_{D7}\mapsto -\tau_{D7}+2\, .
\end{equation}
These are indeed the monodromies found in \cite{Braun:2018fdp} but now reinterpreted as the monodromies around the constituent defects forming a perturbative D7 brane. In the decompactification limit $\text{Vol}(K3)\rightarrow \infty$ the constituents merge and one recovers the elementary D7 brane in ten flat dimensions (this limit is dual to the degeneration limit of the K3 fiber considered in \cite{Braun:2018fdp} in the geometric context).
\begin{figure}
	\centering
	\includegraphics[keepaspectratio,width=14cm]{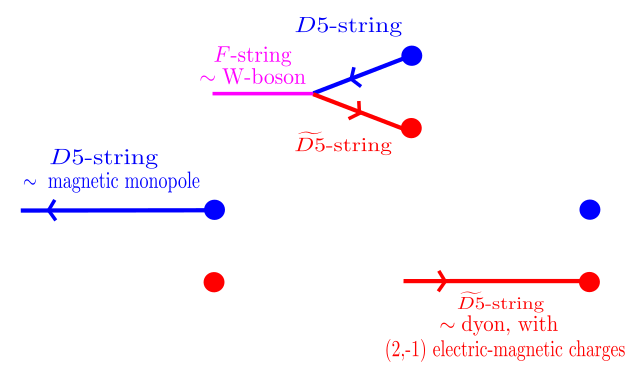}
	\caption{Splitting of perturbative D7 brane into two cosmic defects (red and blue points). A D5-string (blue) can end on one of the defects, and a D5 with a negative unit of electric charge (red) ($\widetilde{\text{D5}}$) can end on the other. The F-string (purple) can end on the entire configuration via a string junction.}
	\label{fig:7-brane-pics1}
\end{figure}

\begin{figure}
	\centering
	\includegraphics[keepaspectratio,width=12cm]{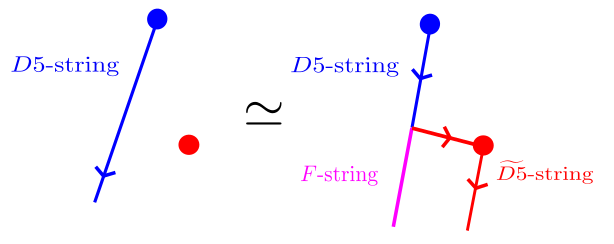}
	\caption{A D5 string ending on the first defect picks up a unit of F-string charge upon half-way encircling a D7-brane. Therefore, after passing an angle $>\pi/2$ one may lower its energy by splitting off an F-string. The remaining $\widetilde{\text{D5}}$ string can end on the second defect.}
	\label{fig:7-brane-pics2}
\end{figure}

Note that $M_1$ and $M_2$ differ from each other only by a choice of base point: we have
\begin{equation}\label{eq:monodromy2}
M_2=T\cdot M_1 \cdot T^{-1}\,  .
\end{equation}
As a D5-string can end on one of the two defects forming a perturbative D7 brane, the string that can end on the other one (let us call it the $\widetilde{\text{D5}}$ string) differs from a D5 string by one unit of F-string charge, see Figure \ref{fig:7-brane-pics1}. Indeed, a D5 string ending on the first defect picks up one unit of F-string charge upon rotating half-way around the transverse plane of a D7 brane due to the non-trivial $C_0$ and $C_4$ profile around it, 
\begin{equation}
C_0\rightarrow C_0-\frac{1}{2}\, ,\quad \int_{K3} C_4\rightarrow \int_{K3} C_4+\frac{1}{2}\, .
\end{equation}
This induces F-string charge on the D5-brane due to its CS-coupling, 
\begin{equation}
S_{CS}\supset 2\pi \int_{\Sigma}B_2 \left(\int_{K3}C_4-C_0\right)\, ,
\end{equation}
see Figure \ref{fig:7-brane-pics2} for a pictorial representation. After splitting off an F-string, which can end on the pair of defects via a string junction, we get the $\widetilde{\text{D5}}$ string that can end on the second defect. Of course there exists an entire tower of monopoles with arbitrary integer electric charge, corresponding to D5 strings winding around the pair of defects, eventually ending on one. By analogy with $(p,q)$-strings in 10d we will refer to $(p',q')$-strings as the strings with charges $(-q',q',p',0)\in\mathbb{Z}^4$ such that the $(1,0)$-string is the F-string and the $(0,1)$-string is the D5-string, and a BPS dyon in the $SU(2)$ Yang-Mills theory with electric-magnetic charges $(2p',q')$ can be thought of as a stretched $(p',q')$-string.

The typical distance separating the two defects is non-perturbative in the 7-brane gauge coupling and thus of order $\Lambda_{\text{eff}} \sim r_{0}e^{\frac{2\pi i}{2} \tau_{D7}(r_0)}$, with arbitrary radial scale $r_0$. Below this distance scale, the logarithmic running of the conformal factor terminates because there is no remaining D3 charge at smaller radii. It is apparent that \textit{this is what keeps the conformal factor from turning negative too close to the location of negative D3-charge}. 

\begin{figure}
	\centering
	\includegraphics[keepaspectratio,width=8cm]{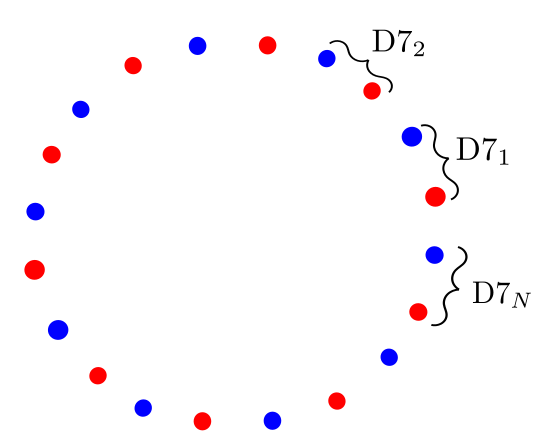}
	\caption{The $2N$ monodromy defects forming $N$ perturbative D7 branes, depicted on the transverse plane, at the origin of the Coulomb branch. Locally, each defect looks the same, but the ones drawn in blue are mutually non-local compared to the ones drawn in red.}
	\label{fig:ND7pic}
\end{figure}

\begin{figure}
	\centering
	\includegraphics[keepaspectratio,width=8cm]{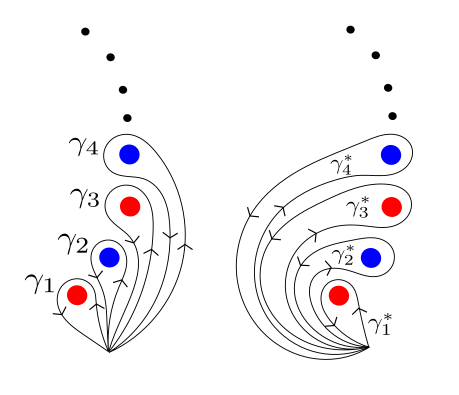}
	\caption{Left: Loops $\gamma_n$ encircling the defects arranged in a circle from the 'outside'. Right: Different choice of loops $\gamma^*_n$ encircling the defects from the 'inside'.}
	\label{fig:monodromy_loops}
\end{figure}

It does not take too much imagination to anticipate what happens in the case of $N$ D7 branes wrapped on K3 (at the origin of the Coulomb branch): non-perturbatively, the instanton corrections to the holographic running of the gauge coupling become important at a distance scale
\begin{equation}
\Lambda\sim r_0e^{\frac{2\pi i}{2N}\tau_{D7}(r_0)}\, ,
\end{equation}
and the $N$ \textit{perturbative} (in the $\alpha'$ expansion) D7-branes split into $N$ pairs of exotic cosmic defects: each pair can be thought of as a bound state realizing a perturbative D7 brane as in eq. \eqref{eq:monodromy1} and \eqref{eq:monodromy2}. At the origin of the Coulomb branch (i.e. $u_k=0$) the characteristic polynomial $\mathcal{W}(x)$ reads
\begin{equation}
\mathcal{W}(x)|_{u_k=0}=x^N\quad \rightarrow \quad \mathcal{W}_{\pm}(x)=x^N\mp \Lambda^N\, ,
\end{equation}
and the entire $\mathbb{Z}_{2N}$ R-symmetry group is unbroken. There,
the $2N$ defects align along a circle of radius $\Lambda$, see Figure \ref{fig:ND7pic},
\begin{equation}
\mathcal{W}_+(x)\mathcal{W}_-(x)=0 \quad \rightarrow x=\Lambda e^{\frac{2\pi i}{2N}k}\, ,\quad  k\in\{0,...,2N-1\}\, .
\end{equation}
Since the R-symmetry of the gauge theory corresponds geometrically to the rotational symmetry around the stack of seven-branes, we see that the non-perturbative splitting of the perturbative D7 branes into $2N$ defects is the bulk version of the explicit symmetry breaking $U(1)_R\rightarrow \mathbb{Z}_{2N}$ by instantons in the gauge theory. The bulk monodromies around the $2N$ defects (with loops $\gamma_{n}$ taken as depicted on the left in Figure \ref{fig:monodromy_loops}) 
take the simple form
\begin{align}\label{eq:monodromyn}
M_{n}=&T^{n-1}\cdot M_1\cdot T^{-(n-1)}\, , 
\end{align}
and with $M_1$ and $T$ as in eq. \eqref{eq:monodromy1} and \eqref{eq:monodromy2}. Thus, the monodromy $M_{n}$ leaves the tension of a D5 string with $-n+1$ units of induced F-string charge invariant.\footnote{Note that while the \textit{tension} of a string that can end on a monodromy defect is invariant under the monodromy transformation, its \textit{charge vector} $\vec{p}\in \mathbb{Z}^4\subset \mathbb{Z}^{26}$ is mapped to $-\vec{p}$ in our case, while strings with \textit{invariant} charge vector are not allowed to end on the defect. Otherwise, there would be unwanted extra massless particle states e.g. from D3 branes wrapped on curves in K3 at singular loci of the gauge theory. The mathematical reason for this in G-theory was pointed out to us by A. Braun: at the location of an elementary monodromy defect, a curve $\gamma$ with self-intersection $\gamma^2=-2$ shrinks in the auxiliary K3 fiber of G-theory. The Picard-Lefschetz formula implies $[\gamma]\rightarrow [\gamma]+\gamma^2 [\gamma]=-[\gamma]$ under the monodromy.} The $2N$ elementary defects look the same \textit{locally} because the monodromies differ from each other only by a choice of base point as is manifest in eq. \eqref{eq:monodromyn}. 

Thus again, the splitting of the perturbative D7-brane into the more elementary monodromy defects halts the running of the coupling at the non-perturbative scale $\Lambda$ and prevents regions of negative conformal factor from appearing. It is also clear that the defects corresponding to the roots of (say) $\mathcal{W}_+(x)$ can be stacked on top of each other by tuning the moduli $u_i$,
\begin{equation}
\mathcal{W}_+(x)=\prod_{i=1}^N(x-e_i^+)\rightarrow x \prod_{i=2}^N(x-e_i^+)\rightarrow ...\rightarrow x^N\, ,
\end{equation}
i.e. they are \textit{mutually local}. The effective strings that can end on them are the D5 strings with minimal induced F-string charge.\footnote{Note however that the dyons from the set of stretched D5 strings between three or more \textit{mutually local} cosmic defects are \textit{non-local} in the sense that the pairwise Dirac-products are generally non-vanishing \cite{Argyres:1994xh,Argyres:1995jj}.}

\begin{figure}
	\centering
	\includegraphics[keepaspectratio,width=10cm]{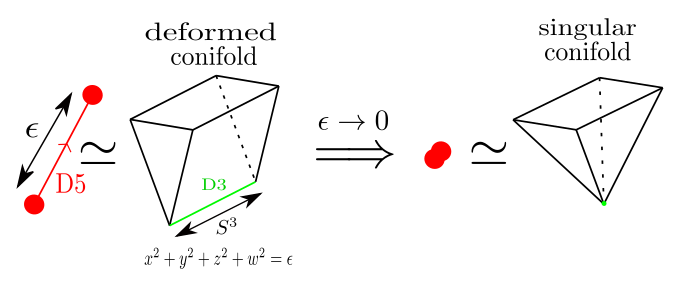}
	\caption{Left: two monodromy defects separated by a small distance $\epsilon$ is dual to the deformed conifold. Right: two defects stacked on top of each other are dual to a singular conifold.}
	\label{fig:two_defects=conifold}
\end{figure}

Correspondingly, the monodromy matrices corresponding to suitably defined loops $\gamma_{n}^*$ (as shown on the right in Figure \ref{fig:monodromy_loops}) around the roots of $\mathcal{W}_+(x)$ satisfy:
\begin{equation}
M_n^*=\begin{cases}
M_1^*\equiv M_1& n\in 2\mathbb{Z}+1\\
M_2^*\equiv M_1\cdot M_2\cdot M_1^{-1} & n\in 2\mathbb{Z}
\end{cases}\, .
\end{equation}
Thus, each perturbative D7 brane splits non-perturbatively into two \textit{mutually non-local} defects (that look the same locally due to $M_2=T^{-1}M_1 T$). Let us call them the A-type and B-type defect respectively. All the A-type (B-type) defects are mutually local with respect to each other, so a D5 string that can end on any of the A-type defects can also end on any of the other A-type defects as long as it stretches along the 'interior' region of the circle of defects. Indeed, such configurations give rise to BPS dyons at the origin of the Coulomb branch \cite{Alim:2011kw}. The field theory that lives on a stack of $n$ A-type (or $n$ B-type) defects is the Argyres-Douglas (AD) CFT of type $(A_1, A_{n-1})$ because these theories are well-known to arise when multiple roots of $\mathcal{W}_{+}(x)$ or $\mathcal{W}_-(x)$ coincide \cite{Argyres:1995jj,Eguchi:1996vu,Eguchi:1996ds,Cecotti:2010fi}. Indeed, the most generic singularity is obtained by colliding two defects, and a single dyon becomes massless. At such a point in moduli space the dual K3 fibered CY threefold should develop a conifold singularity because this is the most generic singularity (at finite distance) that arises in its complex structure moduli space, see Figure \ref{fig:two_defects=conifold}. 

\begin{figure}
	\centering
	\includegraphics[keepaspectratio,width=10cm]{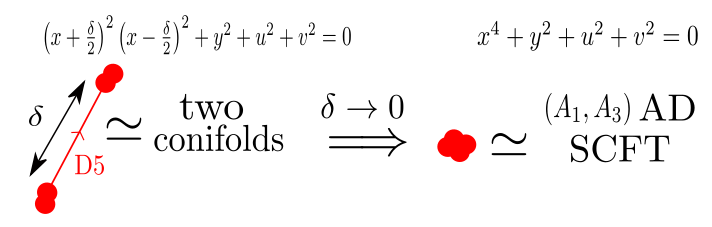}
	\caption{Left: two pairs of two coincident defects, separated from each other a distance $\delta$ corresponds to two conifolds sharing a compact B-cycle with volume $\delta$. Right: colliding the two pairs is dual to shrinking the conifold B-cycle and leads to the geometrically engineered $(A_1,A_3)$ AD CFT.}
	\label{fig:four_defects=AD13}
\end{figure}

For $n\in 2\mathbb{Z}$ we can form $n/2$ pairs of coinciding monodromy defects which is dual to $n$ conifolds on the geometric side. Each conifold hosts a single massless particle from a D3 brane wrapped on the shrinking A-cycle \cite{Strominger:1995cz}, which is dual to the D5 string stretched between two colliding monodromy defects. Furthermore, there exists a basis of $n/2-1$ D5 strings stretching from one pair of defects to the next, and these are dual to D3 branes wrapped on $n/2-1$ compact B-cycles that connect the $n$ conifolds. Stacking all $n$ monodromy defects on top of each other corresponds to colliding the $n$ conifolds, see Figure \ref{fig:four_defects=AD13}. 
Indeed, this is well known to geometrically engineer the AD CFT of type $(A_1, A_{n-1})$ in type IIB string theory \cite{Klemm:1996bj,Katz:1996fh}. Consider for example geometrically engineering the $(A_1,A_3)$ CFT from type IIB on the non-compact CY threefold embedded into $\mathbb{C}^4$ as 
\begin{align}
&f(x)+y^2+u^2+v^2=0\, ,\nonumber\\ \text{with}\quad f(x):=&\left(x+\frac{\delta}{2}\right)^2\left(x-\frac{\delta}{2}\right)^2+\frac{\epsilon_1}{\delta}\left(x+\frac{\delta}{2}\right) +\frac{\epsilon_2}{\delta}\left(x-\frac{\delta}{2}\right)\, .
\end{align}
In the regime $|\epsilon_i|\ll |\delta|^4$ we get two conifolds with deformation parameters $\epsilon_i$ near the loci $\{x=\pm \delta/2,y=u=v=0\}$, and $\delta$ is a measure for the distance between them. In the limit $(\epsilon_i,\delta)\rightarrow 0$ (taken such that $\epsilon_i/\delta\rightarrow 0$) we get $f(x)\rightarrow x^4$ which realizes the unbroken $(A_1,A_3)$ SCFT. The extended Coulomb branch of such SCFT is parametrized by the vev of the CB operator $\langle\mathcal{O}_{\frac{4}{3}}\rangle$, a mass parameter $m$, and a coupling $\mu_{\frac{2}{3}}$, where the subscripts denote conformal dimension. The geometric parameters $(\epsilon_i,\delta)$ correspond to the SCFT parameters as
\begin{equation}
\mu_{\frac{2}{3}} = -\frac{\delta^2}{2}\, ,\quad m = \frac{\epsilon_1+\epsilon_2}{\delta}\, ,\quad \langle \mathcal{O}_{\frac{4}{3}} \rangle = \frac{\delta^4}{16}+\frac{\epsilon_1-\epsilon_2}{2}\, ,
\end{equation}
In other words, the $(A_1,A_1)^2$ point realized by the two conifolds arises at generic points along the locus $\left\{\langle \mathcal{O}_{\frac{4}{3}} \rangle=\frac{1}{4} \mu_{\frac{2}{3}}^2, \  m=0 \right\}$ in the extended Coulomb branch of the $(A_1,A_3)$ theory.

As an aside, it would be very interesting to understand the Higgs branch of the $(A_1, A_{n-1})$ theory with $n$ even in terms of its realization by stacks of cosmic defects in $6d$ supergravity.
Note however that in the fully geometric duality frame one can understand the Higgs branch as a resolution of conifolds. Indeed, as mentioned above, for $n$ even one can find $k:=n/2$ conifolds with $k$ corresponding shrinking $A$-cycle three-spheres connected by $k-1$ compact B-cycles. Using \cite{Candelas:1989ug,Greene:1995hu}, this implies a one-dimensional resolution branch. 
Even more interestingly, it would be important to study the geometrical origin of the flavor symmetry group (isometry group of the Higgs branch) of such theories. The flavor symmetry of $(A_1, A_{n-1})$ AD theories is always $U(1)$ (for $n\geq 6$ and $n$ even) but for $n=4$ it gets enhanced to $SU(2)$.\footnote{This can be clearly seen at least in two ways. Either from the computation of the superconformal index \cite{Maruyoshi:2016aim}, or from the topological symmetry enhancement of their 3d $\mathcal{N}=4$ mirror theory \cite{Xie:2012hs, DelZotto:2014kka}, which is $U(1)$ with $N_f=n/2$ flavors, and it is well known that such enhancement happens only for $N_f=2$.} We leave these interesting questions to further studies.

Finally, we expect that the moduli space of local non-compact CY threefolds dual (via an element in $\mathcal{O}(5,21;\mathbb{Z})$) to a stack of $N$ D7 branes wrapped on K3 can be embedded locally in $\mathbb{C}^4\ni (x,y,u,v)$ via
\begin{equation}
y^2+u^2+v^2=\mathcal{W}(x;\{u_k\})^2-\Lambda^{2N}\, ,
\end{equation}
such that the co-dimension two slice $u=v=0$ is the Seiberg-Witten curve. This non-compact model indeed geometrically engineers the pure $SU(N)$ gauge theory \cite{Klemm:1996bj,Tachikawa:2011yr}.

\begin{figure}
	\centering
	\includegraphics[keepaspectratio,width=10cm]{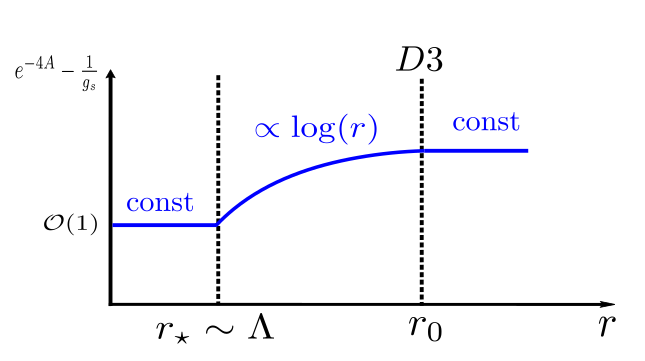}
	\caption{Upon smearing the $2N$ defects forming the $N$ perturbative D7 branes along the angular circle we obtain a simplified effective model, where the conformal factor runs at one-loop in the exterior region $r^*\leq r\leq r_0$ and remains constant in the interior region $r\leq r^*$. Also, we assume that the negative D3 charge on the D7 branes is screened by positive D3 brane charge at $r=r_0$ resulting in a constant profile at $r\geq r_0$ which corresponds to the volume modulus.}
	\label{fig:eff_model}
\end{figure}

Let us distill what we have learned into a rather simple effective model that should be valid when the negative D3 charge is large. We encode the non-perturbative splitting of N D7 branes at the origin of the Coulomb branch as follows: the $2N$ monodromy defects are smeared along a circle of radius $r^* \sim \Lambda$, giving rise to an effective circular defect of \textit{real} co-dimension one that hosts all the negative D3 charge. In this case, the conformal factor runs logarithmically at large radii $r \geq  r^* \sim \Lambda$ and remains constant at smaller radii $r\leq r^*\sim \Lambda$, see Figure \ref{fig:eff_model}. 
There is only one number that this effective description requires as input from a more microscopic analysis: the precise radius $r^*$ at which the effective co-dimension one membrane lives, or equivalently the value of the K3-volume in the interior region $r\leq r^*$. 

Because the interior region can be probed by the individual strong coupling defects forming a perturbative D7 brane 
\begin{figure}
	\centering
	\includegraphics[keepaspectratio,width=7cm]{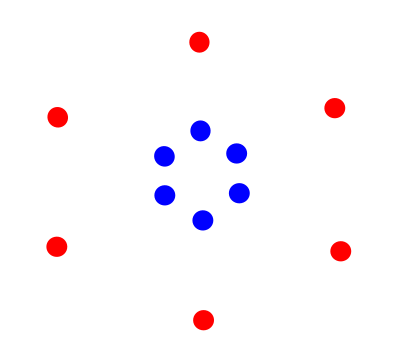}
	\caption{By taking $u_{2,...,N-1}=0$ and $u_N\leq \Lambda^N$ $N$ mutually local exotic defects move into the interior region until they collide at the origin for $u_N=\Lambda^N$ realizing the $(A_1,A_{N-1})$ AD theory.}
	\label{fig:probing_interior}
\end{figure}
(see Figure \ref{fig:probing_interior}) it is natural to propose that the gauge coupling (K3 volume) in the interior region is truly strongly coupled. More precisely, we claim that
\begin{equation}\label{eq:claim}
\tau_{D7}|_{r=0}=\sum_{k=0}^{\infty}\omega_k N^{-k}\, ,
\end{equation}
with numerical $\mathcal{O}(1)$ constants $\omega_k$. In particular, the l.h. side approaches a universal \textit{finite} value $\omega_0$ as $N\rightarrow \infty$. More precisely, we claim that $\omega_0$ cannot be dualized to something weakly coupled, i.e. $0<\text{Im}(\omega_0)<\infty$. In the next section, we will confirm this by inspecting the Seiberg-Witten solution of $SU(N)$ Yang-Mills theory near the origin of the Coulomb branch. The reader not interested in how we arrive at this technical conclusion may skip Section \ref{secsub:resolution_marginal_stability}.

The crucial consequence of our claim is that as the UV-gauge coupling $\text{Im}(\tau_{D7})_{UV}$ is dialed to values $\lesssim N$ the non-perturbatively resolved seven-branes start to explore the entire CY threefold\footnote{From the gauge theory perspective one can think of this as the insertion of R-symmetry breaking spurions with characteristic scale of order the strong coupling scale $\Lambda$. In this sense the 7-brane gauge theory would become truly strongly coupled.}, leaving behind a region with $\mathcal{O}(1)$ Einstein frame curvature. The defects that resolve the apparent negative conformal factor singularity live at real co-dimension two, so we expect to find singularities in the 4d EFT only for isolated values of $\tau_{D7}$ where one of the defects collides with (say) a mobile D3 brane. Therefore, at least when all negative D3 charge comes from a stack of wrapped seven-branes, we expect the 4d effective field theory to generically remain regular even when $\text{Im}(\tau_{D7})_{UV}\ll N$.

\subsection{A strongly curved region}\label{secsub:resolution_marginal_stability}
Now we would like to consider quantitatively the $SU(N)$ gauge theory engineered by D7 branes wrapped on K3 in order to substantiate the claim of eq. \eqref{eq:claim}. To this end, we will analyze the Seiberg Witten solution near the origin of the Coulomb branch $u_{2,...,N}=0$. 

First, at weak coupling $|u_k|\gg 1$ one can relate the bulk field $\tau_{D7}$ to suitable components of the gauge coupling matrix $\tau_{YM}$ of the effective $U(1)^{N-1}$ gauge theory. This goes as follows: the W-bosons and magnetic monopoles correspond to F-strings and D5-strings stretched between the two D7 branes, see Figure \ref{fig:7-brane-pics2}. Thus, for a pair of D7 branes realizing an $SU(2)\subset SU(N)$ Yang-Mills theory its effective $U(1)$ coupling $\tau_{\text{eff}}$ is equal to $\del \mathcal{Z}_M/\del \mathcal{Z}_W$ where $(\mathcal{Z}_M,\mathcal{Z}_W)$ are the central charges of the magnetic monopole and the W-boson \cite{Seiberg:1994rs}. In other words, it measures how fast the mass of a magnetic monopole grows in relation to that of a W-boson upon moving the two D7-branes apart. This identifies $\tau_{\text{eff}}$ with the bulk coupling $\tau_{D7}$ at a point between the two D7-branes.
In the same way, one can reproduce the semi-classical monodromy of the $SU(2)$ Yang-Mills theory from the semi-classical monodromies of the bulk field $\tau_{D7}$ that we discussed in the last section (see Appendix \ref{app:bulk_eom}).
\begin{figure}
	\centering
	\includegraphics[keepaspectratio,width=8cm]{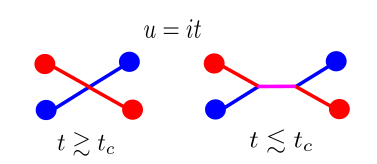}
	\caption{Crossing a wall of marginal stability in the $SU(2)$ theory. Right: weakly coupled chamber where the W-boson is BPS and realized via a string junction system ending on the defects. Left: strongly coupled chamber where the W-boson has decayed into a two-particle state corresponding to the two BPS dyons with electric-magnetic charges $(2,-1)$ and $(0,1)$.}
	\label{fig:SU2_marginal_stability}
\end{figure}

In regions of stronger coupling $|u_k|\gtrsim 1$ this relation breaks down because the F-string actually ends on a string junction as depicted on the r.h. side of Figure \ref{fig:SU2_marginal_stability}. Then, the mass of a W-boson receives a significant contribution from the D5 strings ending in the junction, and we cannot claim $\tau_{D7}\simeq \tau_{\text{eff}}$. At even stronger coupling $|u_k|\lesssim 1$ the W-boson ceases to be BPS and breaks into dyons with electric-magnetic charges $(q_e,q_m)=(2,-1)$ and $(0,1)$, see l.h. side of Figure \ref{fig:SU2_marginal_stability}.\footnote{Note that this does \textit{not} imply that the F-string itself becomes unstable. Rather, the F-string component of the junction system shrinks to zero size.} The co-dimension one surfaces in moduli space across which the BPS property jumps are called \textit{walls of marginal stability} \cite{Seiberg:1994rs,Seiberg:1994aj}. In chambers of moduli space where W-bosons are not BPS, the precise relation between the bulk and Yang-Mills couplings becomes even less clear. 

\begin{figure}
	\centering
	\includegraphics[keepaspectratio,width=10cm]{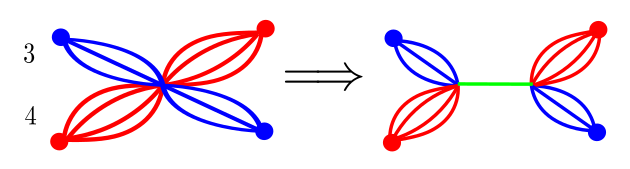}
	\caption{At the wall of marginal stability a BPS multiparticle state with electric-magnetic charge $q'(0,1)+p'(2,-1)$ and $q'/p'>0$ (here depicted for $q=4$ and $p=3$) could reduce its mass by producing a $(p',q'-p')$-string (green) in the middle if this string were parametrically light. But this cannot be the case because $(0,1)$ and $(2,-1)$ are mutually BPS.}
	\label{fig:hypothetical_string}
\end{figure}

Nevertheless, one can use the gauge theory solution to argue for genuine strong bulk coupling $\tau_{D7}$ in the interior region near the origin of the Coulomb branch. First, consider the $SU(2)$ solution of Seiberg and Witten \cite{Seiberg:1994rs}. Along the real imaginary line $u\equiv u_2= i t$ with $t\in \mathbb{R}_+$ the two dyons with charges $(0,1)$ and $(2,-1)$ have equal mass. For $t\geq t_c\approx 0.860$ the W-boson is BPS and at the intersection of the imaginary line with the wall of marginal stability $u=i t_c$ we have $a=a_D$ and thus
\begin{equation}
m_{(0,1)}=m_{(2,-1)}=\frac{1}{2}m_{(2,0)}\, ,
\end{equation}
and all three dyons are \textit{mutually} BPS. At this point in moduli space the W-boson is about to decay into the two dyons (of equal mass), and thus a small F-string localized in the interior must have comparable tension to either of the two D5-strings. We conclude that $\tau_{D7}|_{\text{interior}}=\mathcal{O}(1)$. Finally, let us assume that there is a duality frame in which the bulk coupling is actually weak. Then, there would exist a $(p',q')$ string with parametrically smaller tension than the one of the D5 and F-strings. If moreover $q'/p'>-1$ we could build a corresponding BPS (multi-)particle state in the gauge theory with the same electric-magnetic charges from the dyons with $(0,1)$ and $(2,-1)$. That state could reduce its mass by forming the light string (see Figure \ref{fig:hypothetical_string}). 
This is in contradiction with the fact that $(0,1)$ and $(2,-1)$ are mutually BPS. Thus, such a parametrically light string does not exist. If $q'/p'\leq -1$ the corresponding dyon would not be BPS so excluding light strings in that range is a bit harder, and we will postpone this subtlety until the end of this section when we consider the large-$N$ limit. Ignoring this for now, we conclude that
\begin{equation}
\tau_{D7}|_{\text{interior}}=\mathcal{O}(1)
\end{equation}
in \textit{any} duality frame. In the rest of this section we will show that the origin of the Coulomb branch for $SU(N)$ Yang-Mills theory approaches a wall of marginal stability of the same type in the large $N$ limit $N\rightarrow \infty$. The claim of eq. \eqref{eq:claim} then follows.

\begin{figure}
	\centering
	\includegraphics[keepaspectratio,width=10cm]{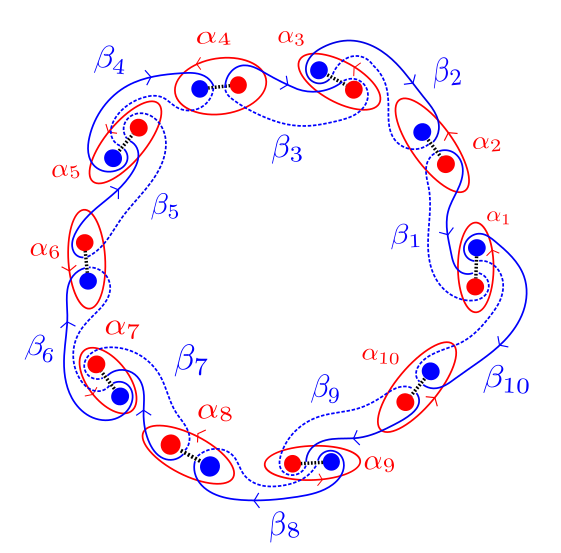}
	\caption{We show (for $N=10$) two sheets of the complex $x$-plane marked by the roots of $\mathcal{W}_+$ (blue) and the roots of $\mathcal{W}_-$ (red), with branch cuts running between them (black dots), at the origin of the Coulomb branch $u_{2,...,N}=0$. The cycles $\alpha_i$ (red) run along the upper sheet, while the cycles $\beta^i$ (blue) run into the branch cuts and continue along the lower sheet (dotted blue).}
	\label{fig:cycle-basis-diagram}
\end{figure}

The Seiberg-Witten solution is determined in terms of the Seiberg-Witten curve of eq. \ref{eq:SWcurve}, and the quantum periods are integrals of the Seiberg-Witten form of eq. \ref{eq:SWform}
over one-cycles of the Riemann surface $\Sigma_{N-1}$. First, let us specify a standard symplectic basis of $H_1(\Sigma_{N-1},\mathbb{Z})$, following \cite{Klemm:1995wp}. As explained in the previous section we identify $\Sigma_{N-1}$ with the double cover of the complex $x$-plane branched over cuts between the $N$ roots $e_+^i$ of $\mathcal{W}_+(x)$ and the $N$ roots $e_-^i$ of $\mathcal{W}_-(x)$, as depicted in Figure \ref{fig:cycle-basis-diagram}.

\begin{figure}
	\centering
	\includegraphics[keepaspectratio,width=8cm]{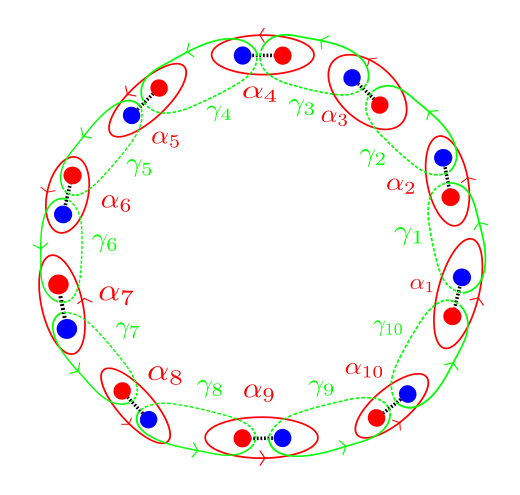}
	\caption{Same setup as in Figure \ref{fig:cycle-basis-diagram} but instead of the $\beta$-cycles we depict the $\gamma$-cycles (in green) which are obtained from the $\alpha$-cycles via action of the R-symmetry.}
	\label{fig:cycle-diagram2}
\end{figure}

Each pair $(e_+^i,e_-^i)$ can be thought of as a perturbative D7 brane wrapped on K3. Let cycles $\alpha_i$, $i=1,..,N$, encircle each pair of roots $(e_+^i,e_-^i)$ counter-clockwise, and define $[\hat{\alpha}_i]:=\sum_{j=1}^i[\alpha_j]$. Furthermore, we let the cycles $\beta^i$, $i=1,...,N-1$, connect the $i$-th and $(i+1)$-th D7 brane as depicted in Figure \ref{fig:cycle-basis-diagram}. The set of cycle classes $\{[\hat{\alpha}_1],...,[\hat{\alpha}_{N-1}],[\beta^1],...,[\beta^{N-1}]\}$ is a symplectic basis of middle homology $H_1(\Sigma_{N-1},\mathbb{Z})$, i.e.
\begin{equation}
\hat{\alpha}_i\cap \beta^j=\delta^j_i\, ,\quad \hat{\alpha}_i\cap\hat{\alpha}_j=\beta^i\cap\beta^j=0\, .
\end{equation}
In order to evaluate the periods at the origin of the Coulomb branch we can use the fact that the $\alpha$-cycles are mapped into each other by discrete rotations in the $x$-plane, i.e. by a $\mathbb{Z}_{N}\subset \mathbb{Z}_{2N}$ R-symmetry group. This implies that the periods $a_l:=\int_{\alpha_l}\lambda$ can be obtained via analytic continuation in the small-$u$ patch as
\begin{equation}\label{eq:analytic_cont1}
a_l(\{u_k\}_k)=e^{\frac{2\pi i}{N}(l-1)}a_1(\{e^{-\frac{2\pi i}{N}k(l-1)}u_k\}_k)\, ,\quad l=1,...,N\, ,
\end{equation}
and $\hat{a}_l=\sum_{m=1}^l a_m$. Similarly, we can obtain the periods $a_D^i$. However, since they are not simply mapped into each other by the R-symmetry group it turns out to be useful to define a further set of cycles $\gamma^{1,...,N}$ drawn in Figure \ref{fig:cycle-diagram2}
that are better adapted to the symmetry of the problem than the $\beta$-cycles. These can be expressed as
\begin{align}
&[\gamma]^l=
-[\beta^i]-(-1)^{l}[\alpha_{l_e+1}] 
\, , \quad  l=1,...,N-1\, ,
\end{align}
where $l_e$ is the even part of $l$. The $\gamma^{i}$ can also be obtained from $\alpha_1$ via the action of the R-symmetry group so the corresponding periods $\pi^{l}_\gamma:=\int_{\gamma^{l}}\lambda$ satisfy
\begin{equation}
\pi_\gamma^{l}(\{u_k\}_k)=e^{\frac{2\pi i}{2N}(2l-1)}a_1(\{e^{-\frac{2\pi i}{2N}k(2l-1)}u_k\}_k)\, ,\quad l=1,...,N\, .
\end{equation}
We can now express the periods $a_D^{l}$ as
\begin{equation}\label{eq:analytic_cont2}
a_D^{l}=-\pi_\gamma^{l}-(-1)^l a_{l_e+1}\, ,\quad l=1,...,N-1\, ,
\end{equation}
and use the continuation formulae \ref{eq:analytic_cont1} and \ref{eq:analytic_cont2} to compute the periods in terms of $a:=a_1$, though we will not explicitly make use of them.

\begin{figure}
	\centering
	\includegraphics[keepaspectratio,width=12cm]{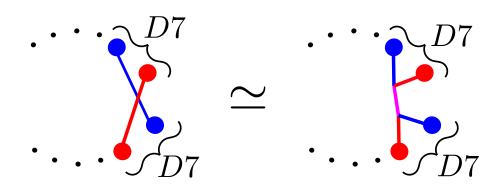}
	\caption{Left: Two D5-strings stretching between neighboring D7 branes forming BPS dyons with central charges $\mathcal{Z}^{d_1}$ and $\mathcal{Z}^{d_2}$. Right: State with central charge $\mathcal{Z}^W=\mathcal{Z}^{d_1}+\mathcal{Z}^{d_2}$ is marginally BPS in the large $N$ limit according to eq. \eqref{eq:Omega}. Upon slightly deforming away from the origin of the Coulomb branch this state should become finitely bound due to the formation of a small F-string (purple).}
	\label{fig:marginal_stability_neighboring}
\end{figure}

\begin{figure}
	\centering
	\includegraphics[keepaspectratio,width=15cm]{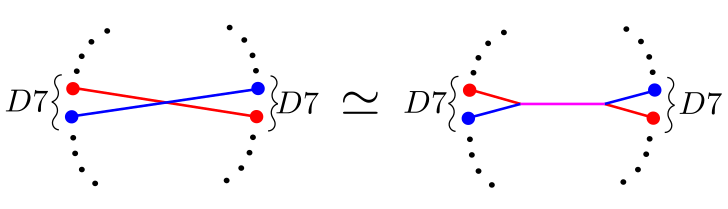}
	\caption{Analogue of Figure \ref{fig:marginal_stability_neighboring} but with pair of D7 branes residing at antipodal positions.}
	\label{fig:marginal_stability_opposite}
\end{figure}

In order to show that the origin of the Coulomb branch approaches a wall of marginal stability as $N\rightarrow \infty$ we consider the ratio of central charges 
\begin{equation}
\Omega(N):=
\frac{\mathcal{Z}^{d_1}}{\mathcal{Z}^W}\, ,
\end{equation}
with $\mathcal{Z}^W=a_{1}-a_{2}$, and $\mathcal{Z}^{d_1}=a_{1}-\pi_\gamma^1$,
as a function of $N$. Using our continuation formulae in eq. \eqref{eq:analytic_cont1} and eq. \eqref{eq:analytic_cont2} it follows that
\begin{equation}
\label{eq:Omega}
\Omega(N)=\frac{1}{1+e^{\frac{2\pi i}{2N}}}=\frac{1}{2}-\frac{i}{2}\frac{\sin\left(\frac{2\pi}{2N}\right)}{1+\cos\left(\frac{2\pi}{2N}\right)}\rightarrow \frac{1}{2}\, \quad \text{as}\quad N\rightarrow \infty\, .
\end{equation}
Thus, in the large $N$ limit the origin of the Coulomb branch of the $SU(N)$ theory is completely analogous to the special point $u=it_c$ on the wall of marginal stability of the $SU(2)$ theory. By reasoning that we applied for the $SU(2)$ theory in the beginning of this section a strong bulk coupling as claimed in eq. \eqref{eq:claim} can be derived: the three relevant (almost-)BPS states are the W-boson with central charge $\mathcal{Z}^W$ and the two dyons (which are exactly BPS \cite{Alim:2011kw}) with central charges $\mathcal{Z}^{d_1}$ and $\mathcal{Z}^{d_2}=-\mathcal{Z}^{d_1}+\mathcal{Z}^W$, depicted in Figure \ref{fig:marginal_stability_neighboring}. 
As a consequence, the bulk gauge coupling $\tau_{D7}$ must take a universal $\mathcal{O}(1)$ value at a point in the interior region close to its boundary, as well as its $2N-1$ images under the R-symmetry group. Even more generally, one can consider almost-BPS W-bosons stretched between the first and $i$-th pair of D7 branes, i.e. with central charges $\mathcal{Z}^W_i=a_{1}-a_{i+1}$, and exactly-BPS stretched dyons with central charges $\mathcal{Z}^{d_1}_i=\sum_{l=1}^i(a_l-\pi^l_\gamma)$, and $\mathcal{Z}^{d_2}_i=-\mathcal{Z}^{d_1}_i+\mathcal{Z}^W_i$, as well as their images under the R-symmetry group to probe essentially the entirety of the interior region (see Figure \ref{fig:marginal_stability_opposite} for the case $i\sim N/2$). 
Again, it is straightforward to show that
\begin{equation}
\Omega_i(N):=\frac{\mathcal{Z}_i^{d1}}{\mathcal{Z}^W_i}\equiv \Omega(N)\, .
\end{equation}
Not only does this confirm that the typical gauge coupling in the interior region is strong, but it also gives further justification for our effective model (cf. Figure \ref{fig:eff_model}) treating the bulk coupling as essentially constant throughout the interior region, at large $N$.

Finally, we can also close the loophole that we left in the beginning of this section: one would like to exclude light $(p',q')$ strings in the interior region also for $q'/p'\leq -1$. If we consider a pair of D7-branes as in Figure \ref{fig:marginal_stability_opposite} the relative separation of two D7-branes diverges in the large-$N$ limit. Then, if we assume that a $(p',q')$ string exists that becomes tensionless in the large-$N$ limit, we can build a dyon from the dyons with central charges $\mathcal{Z}^{d_1}_i$ and $\mathcal{Z}^{d_1}_i$ that can form this light string in the interior as in Figure \ref{fig:hypothetical_string}. The contribution to the mass of this dyon from the junction system is finite in the large $N$ limit\footnote{In units where the BPS dyon connecting neighboring mutually local defects has mass equal to one.} while the contribution to its mass from the light string in the interior grows at most as $N^\alpha$ with $\alpha<1$. Thus, at sufficiently large $N$ the mass of this dyon is smaller than the mass of the BPS dyons we have used as building blocks, so the new dyon should be BPS. But such a state is not part of the BPS-spectrum \cite{Alim:2011kw} so indeed \textit{no} $(p',q')$ string can become tensionless in the interior region as we take the large-$N$ limit.

\section{Comments}
\label{sec:comments}
We have described how the non-perturbative gauge theory effects lead to the splitting of wrapped D7 branes into more elementary cosmic defects and thus prevent regions of negative conformal factor to occur. We have described this rather quantitatively under certain simplifying assumptions: 1) seven-branes wrap a K3 surface, 2) Classical R-symmetry breaking effects from the compact CY as well as other sources of D3 charge can be neglected locally. This means that positively charged D3 branes screen the negative charge on the seven-branes at distances $r_0$ much bigger than the strong coupling scale $\Lambda$.

In this section we would like to sketch plausible outcomes upon dropping our simplifying assumption 2): In \ref{secsub:comments_global_model} we consider type IIB on an $\mathcal{N}=2$ orientifold of $T^2\times K3$ with D3 and D7 branes (the standard weak coupling limit of F-theory on $K3\times K3$ \cite{Sen:1996vd}). We use this model to argue that at small K3-volume $\text{Vol}(K3)\lesssim N$ the D3 charges recombine due to the non-perturbative splitting of seven-branes sweeping up positive D3 charge in the bulk. In \ref{secsub:comments_throat_recombination} we consider the final stages of recombination against the D3 charge hosted by a warped throat. We argue that recombination of D3 charge can be understood as a lowering of the UV cutoff of the KS gauge theory dual the throat. This leads to a reduction of the exponential hierarchy between the IR scale and the UV scale, or in gravitational terms a reduction of warp factor hierarchy. In this extreme regime, the effective field theory of KKLT must be significantly modified (but we will argue in Section \ref{sec:instanton_exp_KKLT} that this regime turns out to be \textit{dynamically} avoided in KKLT).
\subsection{Sketching the small volume limit in a simple global model}\label{secsub:comments_global_model}
We consider the $\mathcal{N}=2$ O7 orientifold of type IIB on $T^2\times K3$ with the orientifold involution acting only on the $T^2$ factor with four fixed points (as analyzed in \cite{Andrianopoli:2003jf}). This is a weak coupling description of F-theory on $K3\times K3$ (as studied in\cite{Bershadsky:1997ec}). There are four O7 planes and 16 D7 branes wrapping $K3$ to cancel the D7-tadpole. Furthermore, we include 24 D3 branes to cancel the D3 tadpole. This allows us to engineer gauge theory sectors with gauge groups of the form $SO(2N)\times USp(2M)$ with $N\leq 16$ D7 branes and $M\leq 24$ D3 branes for D3/D7 branes on O7 planes, and their subgroups $U(N)\times U(M)$ for D3/D7 branes away from O7 planes.

In the unitary case, we can engineer precisely the situation analyzed in the previous section: a stack of $N$ D7 branes surrounded by $N$ D3 branes at radius $r_0$, with $\Lambda\ll r_0$. The classical K3-volume is defined as a modulus at scales $r\geq r_0$. As we take $\Lambda\rightarrow r_0$ limit the D7 branes approach the D3 branes, and the opposite charges start to screen each other. In fact, one can argue that in a sense they actually recombine: from the gauge theory perspective, sending in $\delta N$ D3 branes towards the center means sending the mass of $\delta N$ flavors to zero. This reduces the beta-function coefficient of the seven-brane gauge theory
\begin{equation}
2N\rightarrow 2N-\delta N\, .
\end{equation}
As a consequence, strong coupling effects become important at smaller K3 volume $\text{Vol}(K3)-\frac{1}{g_s}\sim 2N-\delta N$. Moreover, once $\delta N=N$ charge-screening D3 branes have been sent to the origin $r_0\rightarrow 0$ we should instead consider the $U(N)\times U(N)$ gauge theory from $N$ D7 branes and $N$ D3 branes with a massless bi-fundamental hyper and another hypermultiplet in the $(1,\text{Adj})$. The 7-brane stack is still asymptotically free, but this is now only due to the running of the dilaton while the conformal factor remains constant. At a radial scale corresponding to the new strong coupling scale of the 7-brane stack we get $\text{Im}(\hat{\tau})=\text{Im}(\tau)$ and the holomorphic 7-brane coupling vanishes. Below this radial scale the string frame K3 volume is smaller than unity so we should 'T-dualize'\footnote{Note that this is \textit{not} a monodromy transformation.}
\begin{equation}
\hat{\tau}\leftrightarrow \tau\, .
\end{equation}
This corresponds to Seiberg duality on the field theory side \cite{Bershadsky:1996gx}, and sends the $N$ D7 branes and $N$ D3 branes to a stack of $N$ D3 branes. Indeed, for $SU(N_c)$ gauge theory, Seiberg duality makes sense when there are at least as many flavors (D3 branes) as $N_c$ \cite{Seiberg:1994pq}. In the dual description, the dilaton stays constant, and the conformal factor approaches the one of the $AdS_5\times S^5$ throat, with typical length scale of order $N^{\frac{1}{4}}$. 

As we further decrease the K3 volume as defined in the outside region $r\geq r_0$ the radial scale below which one should T-dualize increases further due to the running of the dilaton towards strong coupling at large $r$ until the negative D7 charge of the bulk has been swept up as well. Beyond this, the entire compactification should instead be described by the T-dual frame. In the absence of local charges, i.e. with unbroken (conformal) D7-D3 gauge sector $(SO(8)\times USp(12))^4$, T-duality simply maps the orientifold to itself \cite{Vafa:1997nx}, and the K3-volume is uniquely defined independent of radial scale. For a generic brane configuration it is natural to define the K3-volume and dilaton $\tau$ zero modes $(\hat{\tau}_0,\tau_0)$ at some generic position $z_0$ on the $T^2/\mathbb{Z}_2\simeq \mathbb{P}^1$. As we send $\hat{\tau}_0-\tau_0\rightarrow 0$ the non-perturbatively split 7-branes spread over the $\mathbb{P}^1$ and the spreading of branes should lead to recombination of D3/D7 charges that is complete once we get to the self-dual radius $\text{Im}(\tau_{D7,0})\equiv \text{Im}(\hat{\tau}_0-\tau_0)=0$. 

Therefore, in the range
\begin{equation}
0\leq \text{Im}(\tau_{D7,0})\lesssim 2N\leq 32\, ,
\end{equation}
we expect $\text{Im}(\tau_{D7,0})$ to be a measure of how much un-screened D3/D7 charge remains. We will say more about this once we consider charge recombination against a warped throat in Section \ref{secsub:comments_throat_recombination}. Finally, we expect that the above discussion carries over directly at least for genuine $\mathcal{N}=1$ compactifications on O7 orientifolds with seven-branes wrapped on K3 surfaces. If moreover the CY threefold is K3-\textit{fibered} and the
orientifold involution acts  only on the base of the fibration (with fixed points), one would hope to be able to describe such vacua non-perturbatively in fiber volumes and field-strengths along the fiber by an $\mathcal{N}=1$ version of \cite{Martucci:2012jk,Braun:2013yla,Candelas:2014jma,Candelas:2014kma}.

\subsection{Recombination of D3 charge against a warped throat}
\label{secsub:comments_throat_recombination}
\begin{figure}
	\centering
	\includegraphics[keepaspectratio,width=5cm]{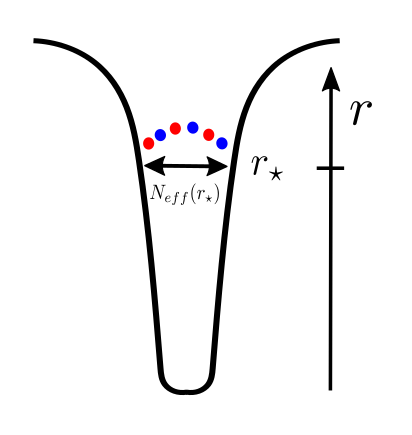}
	\caption{At small values of the overall volume modulus $t\lesssim N_{D3}$ the effective membrane of defects pinches off the throat at a radial scale $r^*$ s.t. $N_{eff}(r^*)\sim t$.}
	\label{fig:throat_recombination}
\end{figure}

In Section \ref{secsub:comments_global_model} we have collected evidence indicating that the non-perturbative resolution of singularities of the conformal factor effectively localizes the negative D3 charge on a real co-dimension one locus that becomes macroscopic as $\text{Vol}(X)^{2/3} \rightarrow |Q_{D3}|$, and sweeps out the entire CY threefold as $\text{Vol}(X)\rightarrow 0$. It appears unavoidable that the amount of positive D3 charge, hosted in fluxes or mobile D3 brane swept up by the membrane effectively recombines with the negative charge on the membrane thus reducing the overall amount of unscreened D3 charge $Q_{D3}$.\footnote{One might expect that such recombination of negative D3 charges on seven-branes against bulk fluxes changes the classical flux superpotential. However, the size of the strongly curved region is exponentially dependent on the vev's of the K\"ahler moduli, so there is no room for additional corrections to the 4d superpotential beyond the standard non-perturbative expansion. We thank Daniel Junghans for a useful discussion about this point.} So far, we have argued for this phenomenon from the perspective of the gauge theories residing on the negatively charged defects, but one can argue for the same result from the perspective of a region carrying overall positive D3 charge:

Specifically, we consider a warped throat with total D3 charge $N_{D3}:=|Q_{D3}|$. As usual, the D3 charge is spread out along a radial direction such that the total integrated D3 charge up to radial distance $r$ is given by
\begin{equation}
N_{eff}(r):=N_{D3}+\frac{3}{2\pi}g_sM^2\log(r/r_{UV})\, ,
\end{equation} 
for $r\leq r_{UV}$ \cite{Klebanov:2000hb,Herzog:2002ih}. From the local solution of the conformal factor, one notices that if this running is cut-off at intermediate radial distance $r_*\ll r_{UV}$ due to e.g. the spreading of seven-branes down into the throat, the apparent volume of the throat at the cut-off value is still of order
\begin{equation}\label{eq:throat_volume}
t\sim \left(\text{Vol}(\text{throat})(r_*)\right)^{\frac{2}{3}}\approx \left(\frac{16}{27}\right)^{\frac{2}{3}}\frac{N_{eff}(r^*)}{4\pi}\, ,
\end{equation}
and the warp factor log-hierarchy between that point and the infra-red end is given by
\begin{equation}
A(r_{IR})-A(r_{UV})=: \log(a_{0}^{-1})\approx \frac{2\pi}{3}\frac{N_{eff}(r^*)}{g_sM^2}\, .
\end{equation}
Since the overall volume modulus corresponds to an additive constant to $e^{-4A}$ we see that for sufficiently small value of the K\"ahler modulus the semi-classical singularity at the locus where $e^{-4A}=0$ starts to walk down the throat, effectively pinching of the throat at smaller radii. At least when the negative charge comes from seven-branes wrapping K3 we can argue that the pinching of the throat is non-perturbatively replaced by a set of elementary monodromy defects, see Figure \ref{fig:throat_recombination}. 
In this regime, one can think of the system as a Randall-Sundrum throat with total D3 charge $N_{eff}(r^*)$ and a Planck-brane residing at $r=r^*$ carrying $-N_{eff}(r^*)$ units of charge. If this occurs at radial position $r^*$, the typical length scale of the setup is given by eq. \eqref{eq:throat_volume}, and we can think of all the positive D3 charge residing at larger radii as having been absorbed by the effective membrane carrying the negative charge.

As a consequence, the throat hierarchy starts to become exponentially sensitive to the value of the K\"ahler modulus due to the screening of charges above the dynamically adjusting UV-cutoff $r^*$,
\begin{equation}
\log(a_0^{-2}) \longrightarrow \frac{4\pi}{3}\frac{N_{eff}(r^*)}{g_sM^2}\equiv 2\pi \frac{\widetilde{\text{Vol}(K3)}}{g_s M^2/c}\, ,\quad c=\mathcal{O}(1)\, .
\end{equation}
Here, the $\mathcal{O}(1)$ constant depends on the precise relation between the throat volume measured near the radial scale $r^*$ and the value of the real part of the holomorphic modulus $\hat{\tau}-\frac{\chi(S)}{24}\tau\simeq \widetilde{\text{Vol}(K3)}$.\footnote{If they were exactly the same we would get
$c=\frac{8\pi}{3} \left(\frac{16}{27}\right)^{-\frac{2}{3}}$\, .
} Crucially, the validity of the standard KKLT EFT starts to break down once a significant fraction of the throat fluxes that induce the running of the warp factor are swept up by the effective membrane carrying the negative D3 charge. This is because the uplift starts to become \textit{exponentially} sensitive to the volume modulus $t$. As argued in \cite{Carta:2019rhx,Gao:2020xqh} this regime begins once $t\lesssim N_{D3}$. However, and this will become crucial in Section \ref{sec:instanton_exp_KKLT} when we discuss the KKLT uplift in more detail, in the regime $t\lesssim N_{D3}$ we also no longer have a relation $S_{ED3}\sim  t$ where $S_{ED3}$ is the action of the leading ED3 instanton contributing to the superpotential.

\section{Resolution of singularities and instanton expansion in KKLT}
\label{sec:instanton_exp_KKLT}
\begin{figure}
	\centering
	\includegraphics[keepaspectratio,width=8cm]{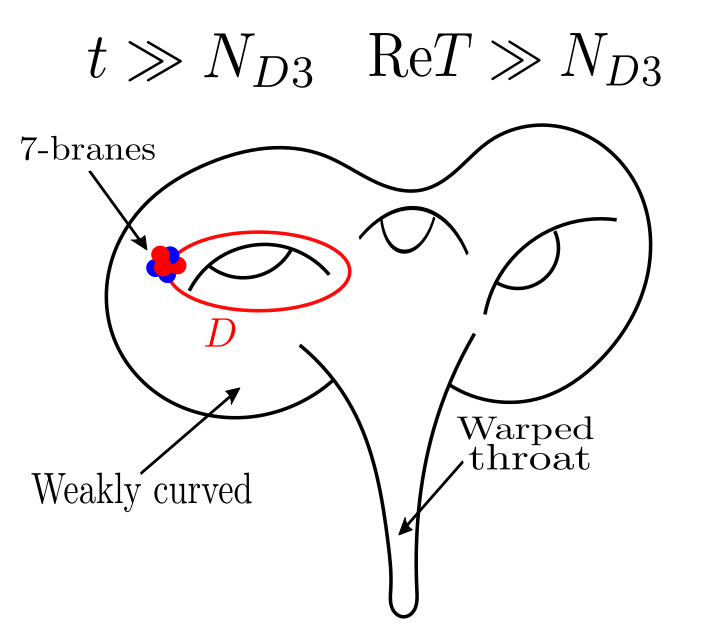}
	\caption{Compact CY with warped throat, seven-branes contributing the D3 tadpole and transversally intersecting rigid divisor $D$ supporting an ED3 instanton responsible for moduli stabilization. We depict the large volume regime where the D7-branes can be treated semi-classically, i.e. they are all stacked on top of each other. The bulk CY is weakly curved everywhere except exponentially close to the seven-branes.}
	\label{fig:bulkCY_LV}
\end{figure}

Now consider again the KKLT scenario and again for simplicity we set $h^{1,1}_+=1$ and assume that the K\"ahler modulus $T$ is stabilized by an ED3 instanton wrapping some divisor $D$. Furthermore, in order for the ED3 instanton to contribute to the superpotential even in the absence of worldvolume and background fluxes, we will assume that $D$ intersects the O7 plane transversally along some curve and that $D$ is rigid, i.e. $h^{i}(\mathcal{O}_D,D)=0$ for $i=1,2$. To simplify things even further we will also place four D7 branes on the O7 plane such that the dilaton is a constant (see Figure \ref{fig:bulkCY_LV}), 
and to make concrete contact with the preceding analysis we will take the divisor wrapped by the O7 plane to be K3. As before, $Q$ denotes the negative D3 charge spread along the seven-branes.

We will also (for now) set the K\"ahler modulus $T$ to a (very) large value. As shown in Section \ref{sec:classical-singularities} in the vicinity of the seven-branes the conformal factor will vary logarithmically, matching the one-loop running of the gauge theory living on the seven-brane stack, while away from the sources it can be treated as a constant. The (complexified) ED3 action is
\begin{equation}
S_{ED3}=2\pi\int_D \left(e^{-4A}\frac{1}{2}J\wedge J-iC_4\right)\, .
\end{equation}
If we assume that $D$ is the generator of the cone of effective divisors this is equal to the real part of our K\"ahler modulus $T$. We may write the unit volume K\"ahler form $J$ as
\begin{equation}
J=\left(\frac{6}{\kappa}\right)^{\frac{1}{3}}[D]\, ,
\end{equation}
with triple intersection number $\kappa$, where $[D]$ is the (harmonic) Poincar\'e dual two-form to the divisor $D$. Using that $dJ=0$ we can relate the divisor volume $\text{Re}(T)$ with the overall volume modulus $t$ (defined in eq. \ref{eq:overall_vol_modulus}),
\begin{align}\label{eq:Kahler_modulus_volume_relation}
\text{Re}(T)&=\int_D e^{-4A}\frac{1}{2}J\wedge J=\int_X e^{-4A}\frac{1}{2}J\wedge J\wedge \delta_D^{(2)}\nonumber\\
&=\left(\frac{9}{2}\kappa\right)^{\frac{1}{3}} t-\frac{1}{2}\int_X d(e^{-4A})\wedge \omega_1\wedge J\wedge J\, ,
\end{align}
with $\delta$-function two-form $\delta_D^{(2)}=[D]+d\omega_1$, for some one-form $\omega_1$, and we have integrated by parts and used that $[D]\wedge [D]\wedge [D]=\kappa \sqrt{g_{CY}}d^6y$.

\begin{figure}
	\centering
	\includegraphics[keepaspectratio,width=8cm]{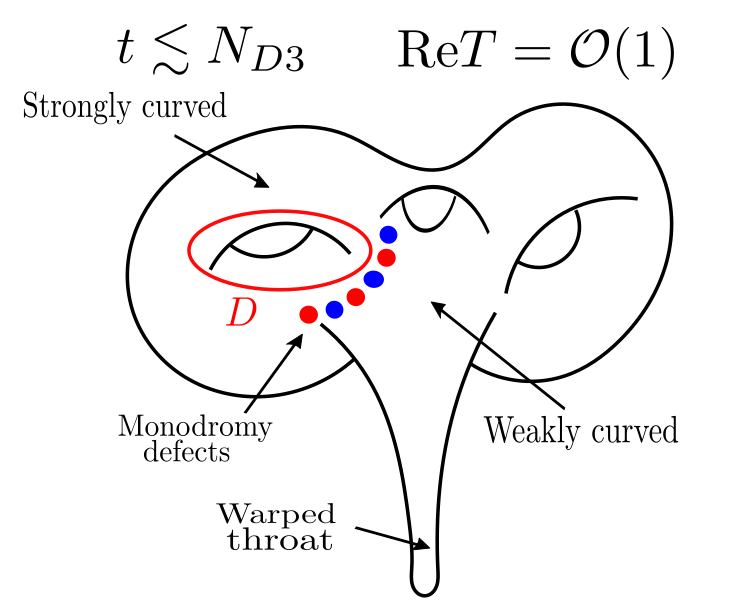}
	\caption{Same CY as in Figure \ref{fig:bulkCY_LV}, for small value of the volume modulus $t\lesssim N_{D3}$ and $\text{Re}\,T=\mathcal{O}(1)$. The non-perturbative splitting of seven-branes has eaten up almost the entire bulk CY but has not yet entered the warped throat. The inside region (on the left) that has been swept up by the spreading of seven-branes is strongly curved, while the outside region (on the right) is still weakly curved.}
	\label{fig:bulkCY_smallV}	
\end{figure}

\begin{figure}
	\centering
	\includegraphics[keepaspectratio,width=8cm]{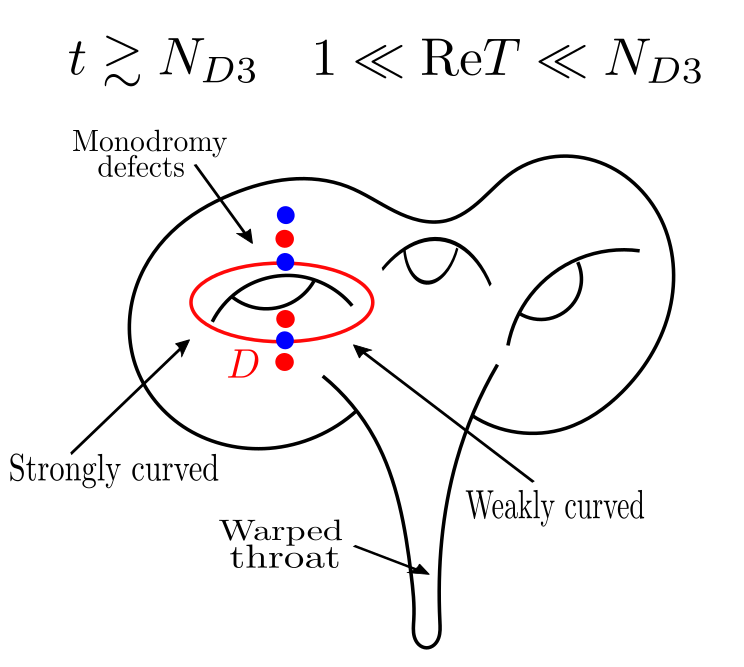}
	\caption{Same CY as in Figure \ref{fig:bulkCY_LV}, for intermediate overall volume $t\gtrsim N_{D3}$ and $1\ll\mbox{Re}\ T \ll N_{D3}$. The non-perturbative splitting of seven-branes has eaten up an $\mathcal{O}(1)$ fraction of the bulk CY but leaves enough room for a large physical volume of the divisor $D$, i.e. Vol$(D)\sim \log(|W_0|^{-1})\gg 1$.}
	\label{fig:bulkCY_intermediateV}	
\end{figure}

In the large volume limit we can treat $e^{-4A}$ as a constant and we get a simple relation between the overall volume modulus and $\text{Re}(T)$
\begin{equation}\label{eq:large-volume-relation}
\text{Re}(T)\rightarrow \left(\frac{9}{2}\kappa\right)^{\frac{1}{3}} t \quad \text{as} \quad t\rightarrow \infty\, ,
\end{equation}
but this relation breaks down once warping becomes significant because the second term in \eqref{eq:Kahler_modulus_volume_relation} cannot be neglected. In particular, in a situation where we have engineered a warped throat carrying a large amount of positive D3 charge $N_{D3}$, we can decrease the overall volume modulus into the critical regime $t\lesssim N_{D3}$ where the classical vanishing locus of the conformal factor $e^{-4A}$ has swept up almost the entire bulk CY but has not yet crept into the warped throat (see Figure \ref{fig:bulkCY_smallV}). 
If the divisor $D$ does not significantly reach into the throat, it has been swept up by the vanishing locus of the conformal factor as well. In particular, as we have argued that volumes are $\mathcal{O}(1)$ in the interior region, it follows that at this point in moduli space the ED3 action is of order one, i.e.
\begin{equation}\label{eq:ReT=O1}
\text{Re}(T)=\mathcal{O}(1)\, .
\end{equation}
Nevertheless, as argued in \cite{Carta:2019rhx,Gao:2020xqh} the overall volume modulus $t$ is of order $N_{D3}$. This is consistent with eq. \eqref{eq:ReT=O1} because the relation of eq. \eqref{eq:large-volume-relation} holds only at very large volume $t\gg N_{D3}$. Since the leading contribution to the non-perturbative superpotential scales as
\begin{equation}
W_{np}\sim e^{-2\pi T}=\mathcal{O}(1)\, ,
\end{equation}
we see that the instanton expansion is poorly controlled at this point in moduli space, and the scale of the stabilizing potential becomes large. Since the warped throat still carries significant D3 charge of order $N_{D3}$, an anti-brane uplift gives a contribution to the scalar potential that is much smaller than that of the non-perturbative superpotential. In other words, the uplift is too small to reach a de Sitter vacuum. But as argued in \cite{Carta:2019rhx}, in the regime where the singularities of the conformal factor can be completely neglected, i.e. $t\gg N_{D3}$, the uplift is too large giving rise to a runaway towards large volume. Hence, by continuity, there must exist a critical value of $\text{Re}(T)$ in the intermediate regime
\begin{equation}
t\gtrsim N_{D3}\, ,\quad 1<\text{Re}(T)< N_{D3}
\end{equation}
such that the uplift precisely competes with the leading small non-perturbative contributions to the scalar potential (depicted in Figure \ref{fig:bulkCY_intermediateV}). 
Thus, in order to promote this point in field space to a de Sitter minimum of the scalar potential it should suffice to find an appropriate small value of $W_0$ such that at the above critical value of $\text{Re}(T)$ one \textit{also} finds
\begin{equation}\label{eq:KKLTmin}
e^{-2\pi \text{Re}(T)}\sim |W_0|\ll 1\, .
\end{equation}
Indeed, the AdS KKLT minimum arises via a competition of the classical flux superpotential with the leading non-perturbative correction and is thus very robust against corrections to the K\"ahler potential, as already pointed out in \cite{Kachru:2003aw}. Even if significant non-perturbative corrections would render the K\"ahler potential hard to compute it is difficult to imagine how this could drive the KKLT minimum to strong coupling $T=T_s=\mathcal{O}(1)$. This would require a quite remarkable fine-tuning $|\del_T K-2\pi|_{T=T_s}\lesssim |W_0|\ll 1$ that string theory would need to automatically realize, given a small $W_0$.\footnote{We thank Daniel Junghans for a useful discussion about this point.}

Interestingly, even though the non-perturbative  expansion of the \textit{superpotential} is controlled by virtue of a small classical flux superpotential, the gauge theory hosted on the seven-branes that dominate the D3 tadpole is strongly coupled in the sense that classical R-symmetry breaking spurions take values of order the strong coupling scale of the gauge theory.
Since an $\mathcal{O}(1)$ fraction of the bulk CY is now strongly curved, one might be worried about the stability of the complex structure since their potential is (in practice) computed using the large volume approximation. However, since the size of the strongly curved region depends explicitly on the K\"ahler moduli, its presence can enter the superpotential only non-perturbatively in $T$. Since a small $W_0$ ensures that non-perturbative corrections are negligible we can still rely on the classical approximation of $W$ in order to compute the F-term potential for the CS moduli and the dilaton.  Again, obtaining a good approximation of the \textit{K\"ahler potential} may be difficult. However, if all moduli have a steep potential at large volume, and if $W_0$ is small, the F-terms are again independent of the K\"ahler potential to good approximation, $ D_aW=\del_aW+\del_a K W_0\approx \del_a W$. If some of the CS moduli have masses of order $W_0$ one has to be a bit more careful, but at least the subset of solutions of the supergravity F-terms of the light moduli which can be deformed to nearby solutions of $\del_aW=0$ should survive because they again rely on a competition between different terms in the superpotential (the solutions of \cite{Demirtas:2019sip,Demirtas:2020ffz,Blumenhagen:2020ire} are of this type). As a consequence we do not see how the presence of the large strongly curved region could jeopardize the scheme of moduli stabilization. Finally, in order to argue for an actual dS vacuum one has to also ensure that the anti-brane uplift potential can be computed reliably. But up to the irrelevant overall $e^K$ factor in the F-term potential, the uplift potential is determined by the local Physics of the IR region of the throat which remains in its supergravity regime near the AdS KKLT minimum. Thus, it appears that even the KKLT dS solution is generically safe from corrections induced by the strongly curved region in the bulk.

Nevertheless, the fact that generically an $\mathcal{O}(1)$ fraction of the bulk CY is strongly curved in KKLT should be kept in mind in the construction of phenomenological models based on the KKLT scenario. Finally, since the local divisor volume (of K3) is not single-valued upon encircling the elementary defects that we have described, we see that at the KKLT minimum a significant part of the full bulk CY may be best thought of as a $U$-fold or $T$-fold (see e.g. \cite{Plauschinn:2018wbo} for a review of non-geometric backgrounds in string theory).\footnote{We thank Alexander Westphal for a helpful discussion about this.}

\section{Conclusions}
\label{sec:conclusions}

In this paper we have revisited a potential control problem for the KKLT scenario of moduli stabilization \cite{Kachru:2003aw}: as argued in \cite{Carta:2019rhx,Gao:2020xqh}, when K\"ahler moduli take values near a meta-stable KKLT de Sitter vacuum the backreaction radii associated with D3 charges from fluxes and branes can no longer be neglected \textit{anywhere} in the compactification. In particular, the singular near-brane behavior of the supergravity fields in the vicinity of sources of \textit{negative} D3 charge extends over an $\mathcal{O}(1)$ fraction of the bulk CY.

We have restricted ourselves to compactifications where negative D3 charge is sourced by seven-branes wrapped on K3 surfaces, realizing (at low energies) $\mathcal{N}=2$ pure Yang-Mills sectors. In Section \ref{sec:resolution} we have argued that the singular near-brane behavior of supergravity fields is resolved by non-perturbative effects in the $\alpha'$-expansion that can be understood quantitatively in terms of instanton effects in the gauge theory. In a way that is analogous to the resolution of dilaton singularities near perturbative (in $g_s$) O7 planes \cite{Sen:1996vd,Banks:1996nj} the non-perturbative corrections in the Seiberg-Witten solution of the IR-behavior of the gauge theory can be understood to \textit{split} a perturbative (in $\alpha'$) D7 brane into a bound state of two mutually non-local defects collectively carrying the negative D3 charge. This non-perturbative splitting of branes then stops the perturbative running of supergravity fields \textit{before} a singularity forms. Concretely, $N$ wrapped D7 branes split into $2N$ defects distributed along a circle around the perturbative location of the branes, leaving behind an 'inside region' with $\mathcal{O}(1)$ local K3-volume (in Einstein frame). Each of the $2N$ defects can be thought of as a cosmic defect in $6d$ $\mathcal{N}=(2,0)$ supergravity, realized by compactifying type IIB string theory on K3. We have described the monodromies of the conformal factor around each defect and by comparing with the singularities in the moduli space of $SU(N)$ Yang-Mills theory we have concluded that a stack of $n$ mutually-local such defects hosts the Argyres-Douglas (AD) SCFT $(A_1,A_{n-1})$.

Because the resolution of singularities can be described entirely in terms of the low energy degrees of freedom already known to be present in the 4d EFT we have concluded that even a \textit{macroscopic} strongly curved 'inside region' would not invalidate the KKLT EFT employed in \cite{Kachru:2003aw} to discuss the stabilization of K\"ahler moduli. Moreover, we have argued in Section \ref{sec:instanton_exp_KKLT} that a small flux superpotential should ensure that a sufficiently large part of the bulk CY remains weakly curved such that the leading ED3 instanton has large action, thus \textit{dynamically} preventing significant alterations of the 4d EFT treatment of the anti-brane uplift.

It would be interesting to understand in detail the resolution of the singularity of the conformal factor also for other sources of negative D3 charge. Arguably, O3 planes are easiest to understand because the singularity of the conformal factor can be related via T-duality to the dilaton singularity of an O7 plane wrapped on $T^4$ which is resolved in F-theory \cite{Sen:1996vd}. Also, it appears plausible that one can understand the case of seven-branes wrapped on rigid divisors because they realize $\mathcal{N}=1$ pure Yang-Mills theories which can be thought of as massive deformations of the $\mathcal{N}=2$ theory that we have considered. Perhaps the most mysterious case is that of wrapped seven-branes on surfaces with many normal bundle deformations. The gauge theory is IR-free so explaining the resolution of the near-brane singularity in terms of gauge theory instantons seems difficult. Furthermore, it would be interesting to turn things around and ask what other AD SCFTs can be realized by stacks of exotic branes in $\mathcal{N}=(2,0)$ supergravity. Finally, one might be able to use our results to improve control over the K\"ahler uplifting scenario of \cite{Westphal:2006tn} where a confining seven-brane theory is at most marginally weakly coupled at the UV-cutoff.\footnote{We thank Alexander Westphal for useful discussions about this.}

\vspace{0.5cm}

\noindent \textbf{Acknowledgments:} We thank Michael Douglas, Daniel Junghans, Manki Kim, Liam McAllister, Thomas Van Riet, Alexander Westphal, Raffaele Savelli and Eva Silverstein for useful discussions and/or comments on a draft. We are particularly grateful to Andreas Braun for explaining ref. \cite{Braun:2018fdp} to us and for useful discussions surrounding it. The work of J.M.~was supported by the Simons Foundation Origins of the Universe Initiative. F.C. is supported by STFC consolidated grant ST/T000708/1.

%\newpage
\appendix
\section{Bulk equations of motion}\label{app:bulk_eom}
We would like to show that the non-perturbatively corrected 10d metric still solves the tree level equations of motion (but with different singular source terms). We will zoom-in close to the stack of D7 branes so we approximate the total space as $\mathbb{C}\times K3$. First, let us forget about D3 brane charges. It is well known that the inclusion of 7-brane charges modifies the Ricci flat 10d metric as (see e.g. \cite{Braun:2008ua})
\begin{equation}
ds^2\rightarrow dx^2+\text{Im}(\tau(a_b))|da_b|^2+ds^2_{K3}\, ,
\end{equation}
where $a_b$ is a complex coordinate parameterizing the transverse complex plane.
As in \cite{Giddings:2001yu}, the inclusion of D3 charges further modifies the metric as
\begin{equation}
ds^2\rightarrow e^{2A(a,\bar{a})}dx^2+e^{-2A(a_b,\bar{a_b})}(\text{Im}(\tau(a_b))|da_b|^2+ds^2_{K3})\, ,
\end{equation}
where $e^{-4A}$ satisfies an electro-static equation
\begin{equation}
-\nabla^2_0\, e^{-4A}=\rho_{D3}\, ,
\end{equation}
with $\rho_{D3}$ the D3-charge density and $\nabla^2_0$ the 6d Laplacian of $\text{Im}(\tau(a_b))|da_b|^2+ds^2_{K3}$. Here, we have normalized $ds^2_{K3}$ to be the unit volume metric on K3. In the approximation where all D3-charges are smeared over K3 the 6d electro-static problem reduces to a two-dimensional one over the transverse space parameterized by $a_b$. Since the conformal factor $\text{Im}(\tau)$ drops out in two dimensions, away from sources the equation of motion reduces to $\del_{a_b} \bar{\del}_{\bar{a_b}}e^{-4A}=0$ which is indeed true when $e^{-4A(a_b,\bar{a_b})}=\text{Im}(\hat{\tau}(a_b))$ for $\hat{\tau}(a_b)$ holomorphic. 

At leading order in the $\alpha'$ expansions, $\tau$ and $\hat{\tau}$ are indeed the gauge couplings on probe D3 and D7 branes as is seen from expanding the DBI action to leading order in the brane velocities,
\begin{align}\label{eq:probeD3-brane-kinetic}
S_{D3}&\supset-2\pi \int d^4x\left(\frac{1}{2}\text{Im}(\tau)|\del a_{D3}|^2+\frac{1}{2}g_{ij}^{K3}\del \phi^i \del \phi^j+...\right)\, ,\\
S_{D7}&\supset-2\pi \int d^4x\left(\frac{1}{2}\text{Im}(\hat{\tau})|\del a_{D7}|^2+...\right)\, ,\label{eq:probeD7-brane-kinetic}
\end{align}
where $ds^2_{K3}\equiv g_{ij}^{K3}d\phi^i d\phi^j$ is the unit volume K3 metric, and $a_{D3/D7}$ denotes the position of the D3 respectively D7 brane in the transverse plane. Inspecting the CS term of a D7-brane one sees that the $\alpha'^2$ correction leads to the replacement $\hat{\tau}\rightarrow \tau_{D7}:=\hat{\tau}-\frac{\chi(K3)}{24}\tau$ in \eqref{eq:probeD7-brane-kinetic}. This is indeed what is required by $\mathcal{N}=2$ supersymmetry: the kinetic terms of the scalars in vector multiplets are the imaginary parts of the gauge couplings and the field space metric for the scalars in hypermultiplets is hyperk\"ahler because K3 is hyperk\"ahler.

It remains to be shown that $\text{Re}(\hat{\tau})$ is identified with the axion $\int_{K3}C_4$. For this, we simply plug in the solution for the warp factor into the solutions of \cite{Giddings:2001yu}
\begin{align}
F_5=&(1+*)d e^{4A}\wedge d^4x
=d e^{4A}\wedge d^4x + \omega_{K3}\wedge i(\del_{a_b}-\bar{\del_{a_b}})e^{-4A(a_b,\bar{a_b})}\\
=&d\left(e^{4A}\wedge d^4x+\omega_{K3} \,\text{Re}(\hat{\tau}(a_b))\right)\equiv d C_4\, ,
\end{align}
where $\omega_{K3}$ is the volume form on K3. Here, we have used that $e^{-4A(a_b,\bar{a_b})}=\text{Im}(\hat{\tau}(a_b))$ and $i(\del-\bar{\del})\text{Im}(f)=d \text{Re}(f)$ for a holomorphic function $f$.
So indeed, it follows that $\text{Re}(\hat{\tau})=\int_{K3}C_4$.

Finally, the action of a (energy-minimizing) F-string stretched between two D7 branes at points $(a_b^{(1)},a_b^{(2)})$ is given by
\begin{equation}
S_{F}=-2\pi \int_{\Sigma_2}\frac{d^2\sigma \sqrt{g_{\Sigma_2}}}{\sqrt{\text{Im}(\tau)}}  =-2\pi \int dt \left|\int_{a_b^{(1)}}^{a_b^{(2)}}da \right|=-2\pi|a_b^{(1)}-a_b^{(2)}|\int dt\, ,
\end{equation}
where $\Sigma_2$ denotes the worldsheet and $g_{\Sigma_2}$ is the induced \textit{Einstein frame} metric on the worldsheet.
Therefore, defining $a:=a_b^{(1)}-a_b^{(2)}$ we get the central charge of a W-boson in the $SU(2)$ theory realized by the two D7 branes. Since the semi-classical monodromy in the field theory corresponds to exchanging the two D7 branes we get indeed that
\begin{equation}
a\rightarrow -a\, ,
\end{equation}
as in \cite{Seiberg:1994rs}. Likewise, the tree-level DBI action of a stretched (energy-minimizing) string from a D5 brane wrapped on K3 is given by
\begin{equation}
S_{D5}=-2\pi \int_{\Sigma_6}\frac{d^6\sigma \sqrt{g_{\Sigma_6}}}{\sqrt{\text{Im}(\tau)}}=-2\pi \left|\int da_b \,e^{-4A(a_b,\bar{a_b})}\right|\int dt \, ,
\end{equation}
which is corrected due to induced D1 and F-string charge from the K3-curvature and $\int_{K3}C_4$-profile respectively to give
\begin{equation}
S_{D5}=-2\pi |a_D|\int dt\, ,\quad \text{with}\quad a_D:=\int_{a_b^{(1)}}^{a_b^{(2)}}da_b \,\tau_{D7}\, .
\end{equation}
Under the field theory monodromy at infinity the central charge of the magnetic monopole (as realized by a D5 string stretched between two D7 branes) $a_D$ indeed transforms as
\begin{equation}
a_D\rightarrow -a_D+2a\, ,
\end{equation}
as in \cite{Seiberg:1994rs}. The minus sign again comes from the fact that the two D7s are interchanged, and the additive term is generated because the stretched D5 string is rotated half-way around each of the two D7 branes thus picking up the monodromy of the bulk field $\tau_{D7}$ as described in Section \ref{secsub:resolution_qualitative}.

%\newpage

\bibliography{D7SW}
\bibliographystyle{JHEP}
\end{document}